\documentclass[nohyperref]{article}

\usepackage{microtype}
\usepackage{graphicx}
\usepackage{subfigure}
\usepackage{color}
\usepackage{eqparbox,array} 
\usepackage{algorithm,algorithmic}
\renewcommand\algorithmiccomment[1]{% 
    \hfill\#\ \eqparbox{COMMENT}{#1}% 
} 
 
\renewcommand{\algorithmicrequire}{\textbf{Input:}}
\renewcommand{\algorithmicensure}{\textbf{Output:}}
\usepackage{booktabs} % for professional tables

\usepackage{hyperref}

\usepackage[accepted]{icml2023}
\usepackage{amsmath}
\usepackage{amssymb}
\usepackage{mathtools}
\usepackage{amsthm}
\usepackage{dsfont}
\usepackage{diagbox}
\usepackage{adjustbox}
\usepackage{multirow}
\usepackage{makecell}

\usepackage[capitalize,noabbrev]{cleveref}

\theoremstyle{plain}
\newtheorem{theorem}{Theorem}[section]
\newtheorem{proposition}[theorem]{Proposition}

\theoremstyle{definition}

\theoremstyle{remark}

\usepackage{xspace} % used in \method
\usepackage{paralist} % for compactitem

\newcommand{\red}[1]{\textcolor{red}{#1}}

\newcommand{\blue}[1]{\textcolor{blue}{#1}}
\newcommand{\method}{\textsc{DecompDiff}\xspace}

\usepackage[compact]{titlesec}
\titlespacing*{\section}
{0pt}{1.1ex}{1.1ex}
\titlespacing*{\subsection}
{0pt}{1.1ex}{1.1ex}
\titlespacing*{\subsubsection}
{0pt}{1.1ex}{1.1ex}

\icmltitlerunning{\method: Diffusion Models with Decomposed Priors for Structure-Based Drug Design}

\usepackage{amsmath,amsfonts,bm}

\def\eqref#1{equation~\ref{#1}}
\def\1{\bm{1}}

\def\rvepsilon{{\mathbf{\epsilon}}}

\def\rvb{{\mathbf{b}}}

\def\rve{{\mathbf{e}}}

\def\rvh{{\mathbf{h}}}

\def\rvm{{\mathbf{m}}}

\def\rvv{{\mathbf{v}}}

\def\rvx{{\mathbf{x}}}

\def\rmI{{\mathbf{I}}}

\def\vzero{{\bm{0}}}

\def\vmu{{\bm{\mu}}}

\def\vtheta{{\bm{\theta}}}

\def\vb{{\bm{b}}}

\def\vv{{\bm{v}}}

\def\vx{{\bm{x}}}

\def\mH{{\bm{H}}}
\def\mI{{\bm{I}}}

\def\mSigma{{\bm{\Sigma}}}

\DeclareMathAlphabet{\mathsfit}{\encodingdefault}{\sfdefault}{m}{sl}
\SetMathAlphabet{\mathsfit}{bold}{\encodingdefault}{\sfdefault}{bx}{n}

\def\gC{{\mathcal{C}}}

\def\gM{{\mathcal{M}}}
\def\gN{{\mathcal{N}}}

\def\gP{{\mathcal{P}}}

\def\sO{{\mathbb{O}}}

\newcommand{\R}{\mathbb{R}}

\begin{document}

\twocolumn[
% \icmltitle{DecompDiff: Equivariant Diffusion Models for Pocket-based
% Molecule Generation with Decomposable Priors}
\icmltitle{\method: Diffusion Models with Decomposed Priors\\ for Structure-Based Drug Design}



\icmlsetsymbol{equal}{*}

\begin{icmlauthorlist}
\icmlauthor{Jiaqi Guan}{equal,uiuc,bytedance}
\icmlauthor{Xiangxin Zhou}{equal,ucas,casia,bytedance} 
\icmlauthor{Yuwei Yang}{bytedance}
\icmlauthor{Yu Bao}{bytedance} \\
\icmlauthor{Jian Peng}{uiuc}
\icmlauthor{Jianzhu Ma}{thu}
\icmlauthor{Qiang Liu}{ucas,casia}
\icmlauthor{Liang Wang}{ucas,casia}
\icmlauthor{Quanquan Gu}{bytedance}
\end{icmlauthorlist}

\icmlaffiliation{uiuc}{Department of Computer Science, University of Illinois Urbana-Champaign, USA}
% \icmlaffiliation{cas}{School of Artificial Intelligence, University of Chinese Academy of Sciences. Center for Research on Intelligent Perception and Computing (CRIPAC), State Key Laboratory of Multimodal Artificial Intelligence Systems (MAIS), Institute of Automation, Chinese Academy of Sciences (CASIA)}
\icmlaffiliation{ucas}{School of Artificial Intelligence, University of Chinese Academy of Sciences}
\icmlaffiliation{casia}{Center for Research on Intelligent Perception and Computing (CRIPAC), State Key Laboratory of Multimodal Artificial Intelligence Systems (MAIS), Institute of Automation, Chinese Academy of Sciences (CASIA)}
\icmlaffiliation{bytedance}{ByteDance Research (Work was done during Jiaqi's and Xiangxin's internship at ByteDance)}
\icmlaffiliation{thu}{Institute for AI Industry Research, Tsinghua University, Beijing, China}

\icmlcorrespondingauthor{Xiangxin Zhou}{zhouxiangxin1998@gmail.com}
\icmlcorrespondingauthor{Quanquan Gu}{quanquan.gu@bytedance.com}

\icmlkeywords{Structure-based Drug Design, Diffusion Model}

\vskip 0.3in
]



\printAffiliationsAndNotice{\icmlEqualContribution} % otherwise use the standard text.

\begin{abstract}
Designing 3D ligands within a target binding site is a fundamental task in drug discovery. 
Existing structured-based drug design methods treat all ligand atoms equally, which ignores different roles of atoms in the ligand for drug design and can be less efficient for exploring the large drug-like molecule space.
In this paper, inspired by the convention in pharmaceutical practice, we decompose the ligand molecule into two parts, namely arms and scaffold, and propose a new diffusion model, \method, with decomposed priors over arms and scaffold.
In order to facilitate the decomposed generation and improve the properties of the generated molecules, we incorporate both bond diffusion in the model and additional validity guidance in the sampling phase. 
Extensive experiments on CrossDocked2020 show that our approach achieves state-of-the-art performance in generating high-affinity molecules while maintaining proper molecular properties and conformational stability, with up to $-8.39$ Avg. Vina Dock score and $24.5\%$ Success Rate. The code is provided at \url{https://github.com/bytedance/DecompDiff}
 % \todoq{it is better to use number or percentage to highlight the improvement}
\end{abstract}
\section{Introduction}
\label{sec:intro}

Modern deep learning is revolutionizing many subfields in drug discovery, among which 
structure-based drug design (SBDD) \citep{anderson2003process} is an important yet challenging one. 
Aiming at generating 3D ligand molecules conditioned on a target binding site, SBDD requires models to generate drug-like molecules with stable 3D structures and high binding affinities to the target. 
Recently, deep generative models %\todoq{add references here} 
have been successfully employed to achieve this goal. For example, autoregressive models \citep{luo2021autoregressive,liu2022graphbp,peng2022pocket2mol} have achieved promising performance in SBDD tasks, which generate 3D molecules in the target binding site by adding atoms and bonds iteratively. However, autoregressive models suffer from error accumulation and require a generation order, which is nontrivial for molecular graphs. In order to overcome the limitation of autoregressive models, recent works \citep{guan20233d,schneuing2022structure,lin2022diffbp} use diffusion models \citep{ho2020denoising} to approximate the distribution of atom types and positions from a standard Gaussian prior, and use a post-processing algorithm to assign bonds between atoms. 
These diffusion model-based methods can model local and global interactions between atoms simultaneously and achieve better performance than autoregressive models.
Despite the state-of-the-art performance, %these methods have the following drawbacks 1) generating the entire ligand from a standard Gaussian prior can be inefficient as the 3D atom distribution is multimodal, 2) 
existing diffusion model-based approaches neglect bonds in the modeling process, which may lead to unreasonable molecular structures. Moreover, diffusion model-based approaches treat the ligand molecule as a whole and learn the overall correspondence between the target binding site and the ligand. 
However, atoms within the same ligand can be designed for different functions. Therefore, treating all ligand atoms equally may not be the best way for SBDD, especially considering the tremendous drug-like space \citep{virshup2013stochastic} to explore and the limited amount of high quality target-ligand complexes \citep{berman2000protein} for training.
This motivates us to study how to properly incorporate function-related prior knowledge into diffusion model-based SBDD methods.
% TODO: I cannot select one representative publication for molecule property prediction.
\begin{figure}[t]
\begin{center}
\centerline{\includegraphics[width=\columnwidth]{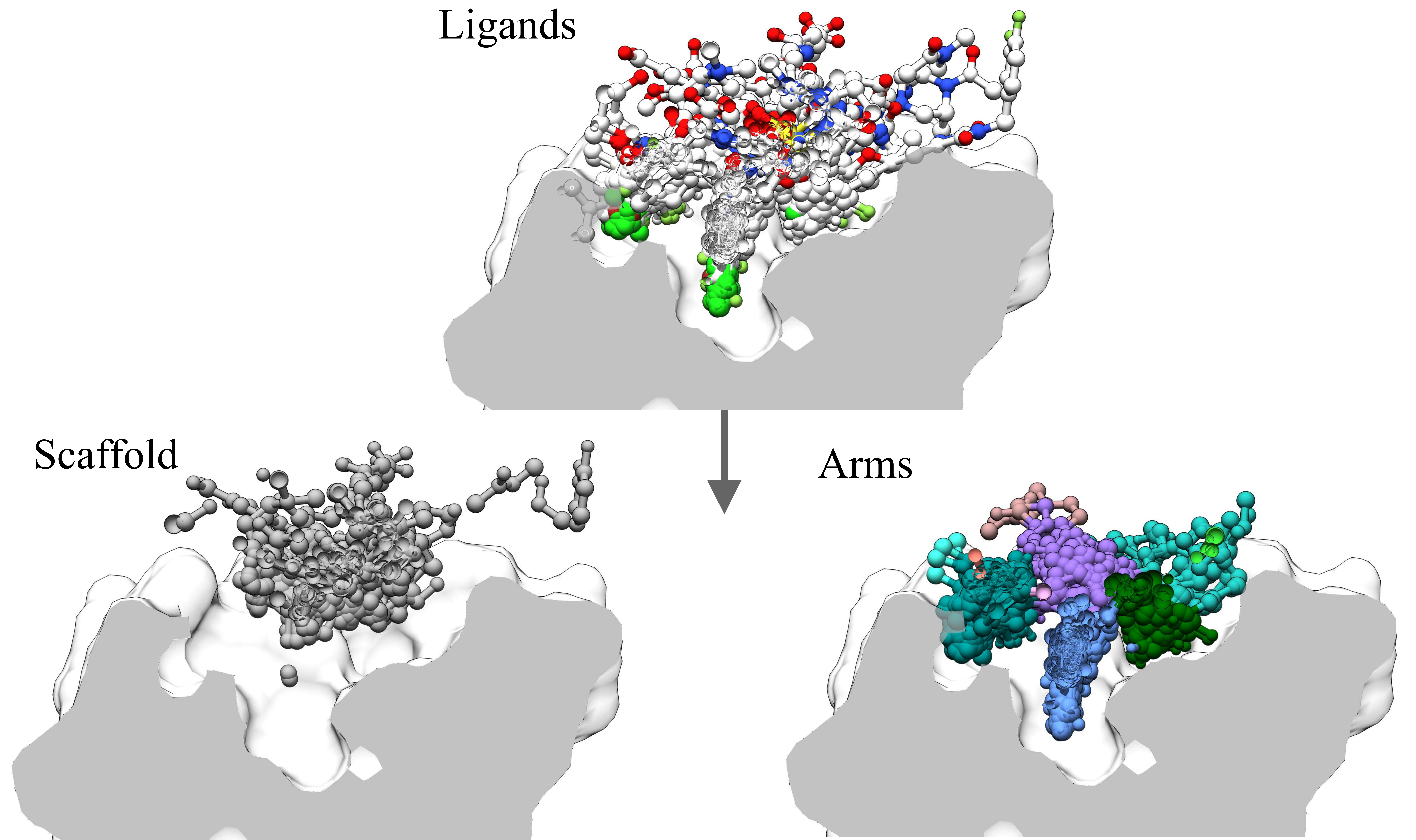}}
\caption{Ligand molecules can be decomposed into arms and scaffold. 
Using MDM2 as an example, small molecule ligands are collected and displayed in the upper panel (colors represent different atom types). 
The ligand atoms are separated into arms and scaffold based on their distance to the protein surface. 
Arms (lower right) form direct contact with the target, while scaffold (lower left) connects the arms together.
Arm atoms are further clustered based on their positions, and the cluster (colored atom groups) show strong shape complementarity with local subpockets.}
\label{fig:decomp}
\vspace{-10mm}
\end{center}
\end{figure}

In fact, decomposing ligands into smaller functional regions is a common practice in conventional drug design. %However, it is underexplored in deep learning-based SBDD methods such as deep generative models. 
As shown in Figure~\ref{fig:decomp}, ligands can be decomposed into scaffold and arms. 
The arms are responsible for interacting with the target to achieve high binding affinity, while the scaffold is responsible for positioning the arms into the desired binding regions. 
In lead optimization, a scaffold is identified first, and then a series of analogs are developed by altering the arms for further activity optimization \citep{wermuth2011practice}; in scaffold hopping practice, arms (or fragments) are placed on the target surface first, then a scaffold (or linker) is placed to connect the arms \citep{schneider1999scaffold}. 
%\citep{imrie2020deep,huang20223dlinker,igashov2022equivariant}
Inspired by the convention in traditional drug design, we aim to incorporate decomposed molecules, i.e., arms and scaffold, into diffusion models. 

% To tackle this problem, separating scaffolds and arms to decompose one large chemical space into the Cartesian product of several smaller chemical spaces is common in practical drug design. 
% In Figure~\ref{fig:decomp}, a ligand can be separated into a scaffold and arms. The arms are responsible for interacting with the receptor. The scaffold is responsible for positioning the arms into favorable interacting regions.
% This idea is also utilized in structure-based virtual screening \citep{sadybekov2022synthon}.
% need to show the idea in antibody design?
% However, the use of scaffold and arm decomposition is still under-explored in pocket-based 3D molecule generation.

%Recently, diffusion models have shown promise in many applications including SBDD, and offer suitable foundation for decomposable molecular generation. 
%Before the prevalence of diffusion models, 
% 3D molecule generation models show superiority to 2D-based models due to awareness of conformation which reflects some important properties of molecules, especially in the scenario of pocket-based molecule generation. 
%Autoregressive models, which hold strong expressive power in theory, are utilized to directly generate molecules in 3D space \citep{luo2021autoregressive,liu2022graphbp,peng2022pocket2mol}. 
%However, imposing a generation order on the molecular graph required by autoregressive models is non-trivial. Autoregressive models suffer from error accumulation and perform worse when generating larger molecules. 

In this paper, we propose \method, a new diffusion model with data-dependent decomposed priors for SBDD.
The decomposed priors respect the natural decomposition of a ligand into arms and scaffold when it interacts with a target. We also introduce a diffusion process on bonds to incorporate the bond generation in an end-to-end fashion instead of post-processing. 
To facilitate the generation process, we develop additional validity guidance in the sampling phase, such as promoting the connection between the scaffold and arms and avoiding a geometric clash between the generated molecule and the target.
% Thus we propose a new diffusion model with data-dependent decomposed prior for pocked-based molecule generation, named DecompDiff, respecting the natural decomposition of a ligand when it interacts with a protein. 
% Besides, we introduce an additional drift term to guide the Langevin sampling towards our preferences, such as promoting the connection between the scaffold and arms and avoiding a geometric clash between the generated molecule and protein.
We highlight our main contributions as follows:
% \begin{itemize}
\begin{compactitem}
\item We propose a diffusion model with decomposed priors for structure-based drug design, which incorporates the natural decomposition of a ligand molecule into function-related regions.
\item We consider both atom and bond diffusion processes in the model to simultaneously generate atoms and bonds for improving drug-likeness and synthesizability.
\item We design and incorporate several guidance terms in the decomposed generation process to improve the molecular validity. 
\item Putting all the above pieces together, our method can generate ligand molecules with a $-8.39$ Avg. Vina Dock score and $24.5\%$ Success Rate,
% better binding affinities with a -8.39 Vina Dock score and a 24.46\% higher success rate, 
achieving the new SOTA on the CrossDocked2020 benchmark. 
% \todoq{it is better to use number or percentage to highlight the improvement}
\end{compactitem}

% with up to $-8.39$ Avg. Vina Dock score and $24.5\%$ Success Rate.

\section{Related Work}
\label{sec:related}

\noindent\textbf{Structure-Based Drug Design\ }SBDD aims to generate 3D molecules in the presence of a target binding site. 
Early attempts use molecular docking for indirect consideration of the target \citep{yang2021knowledge, li2021structure}, which optimize ligand in the 3D space with docking score as a reward. 
Instead of considering every aspect of the target binding site, \citet{long2022zero,adams2022equivariant} proposed to only use the shape information for ligand generation. 
More recent work directly model the correspondence between targets and ligands using target-ligand complexes. 
\citet{ragoza2022generating} represented molecules as atomic density grids and used conditional variational autoencoders to learn the 3D ligand distributions. 
\citet{luo20213d, liu2022generating, peng2022pocket2mol} employed autoregressive models to generate atoms (and bonds) step-by-step. 
%\citet{luo20213d} introduce a mask-fill training schema to model 3D atom occurrences and use an autoregressive model to place atoms in the target binding site. 
% More efforts have been made within the autoregressive framework for SBDD \citep{liu2022generating, peng2022pocket2mol}, and in addition to atoms \citet{peng2022pocket2mol} propose to include bonds in its modeling process. 
Recently, diffusion models start to play a role in SBDD \citep{schneuing2022structure, guan20233d}, and existing approaches denoise atom types and positions sampled from a Gaussian prior.
Unlike the previous approaches, which treat all the ligand atoms as a whole, our method decomposes ligand into arms and scaffold, and incorporate related prior knowledge into diffusion models for better molecular generation.

\noindent\textbf{Decomposed Molecular Generation\ } 
The key idea of decomposed molecular generation is to divide a ligand into subregions and generate each part separately. 
This approach can significantly reduce the search space for drug-like molecules by constraining the search effort in more confined regions. 
Scaffold hopping (or linker design) aims to generate a scaffold/linker to connect the existing arms/fragments. 
\citet{yang2020syntalinker} developed a linker design algorithm to connect two SMILES strings. 
DeLinker \citep{imrie2020deep}, DEVELOP \citep{imrie2021deep} and 3DLinker \citep{huang20223dlinker} extend the application to connect two 3D fragments using autoregressive models.
DiffLinker \citep{igashov2022equivariant} further relieve the constraints on the number of fragments to be connected. 
On the contrary, lead optimization starts from a scaffold structure and optimizes it by adding arms/fragments.
%\citet{jin2018learning, jin2020hierarchical} formulate this task as graph-to-graph translation. This approach learns the transformation between two similar molecules, one of which has better properties.
% Another line of work forces the generated molecules to contain a specific scaffold. 
\citet{arus2020smiles} developed a SMILES-based scaffold decoration method. \citet{lim2020scaffold} and \citet{li2019deepscaffold} achieved this goal on 2D molecular graphs, and \citet{imrie2021deep} took consideration of the 3D scaffold information when generating 2D optimized ligands.  
However, the aforementioned methods all need a reliable initial structure for molecular design. In contrast, our method retains the benefits of molecular decomposition and can be applied to \textit{de novo} drug design to generate molecules from scratch.

\noindent\textbf{Diffusion Models\ }
Diffusion models~\citep{sohl2015deep,ho2020denoising,song2019generative}, which generate samples by iteratively denoising data points sampled from a prior distribution, have shown promising results in generating images \citep{dhariwal2021diffusion,nichol2021glide,ramesh2022hierarchical}, texts~\citep{li2022diffusion}, and speech~\citep{kong2021diffwave}. 
% In its primary settings, the diffusion model is only used to learn data distribution. 
Considering that the initial diffusion model aims to learn the data distribution, it is often necessary to modify it for various realistic controlled generation scenarios.
Some researchers introduce classifier guidance \citep{dhariwal2021diffusion}, or classifier-free guidance \citep{ho2022classifier} for introducing controllable goals.
% Designing informative priors is also helpful for improving diffusion models. 
% Here, we review such topics in conditional diffusion models.
\citet{lee2021priorgrad} leveraged conditional information as a non-standard data-dependent adaptive prior for improving conditional denoising diffusion models.
\citet{vignac2022digress} further showed that the prior distribution closer to data distribution can lead to superior performance.
These works strongly support our design of the prior distribution for the diffusion model. 
It is worth noting that we design priors for more complex 3D structure data, which is less studied in previous work. 

\section{Method}
\label{sec:method}

\begin{figure*}[ht]
\begin{center}
\centerline{\includegraphics[width=0.9\textwidth]{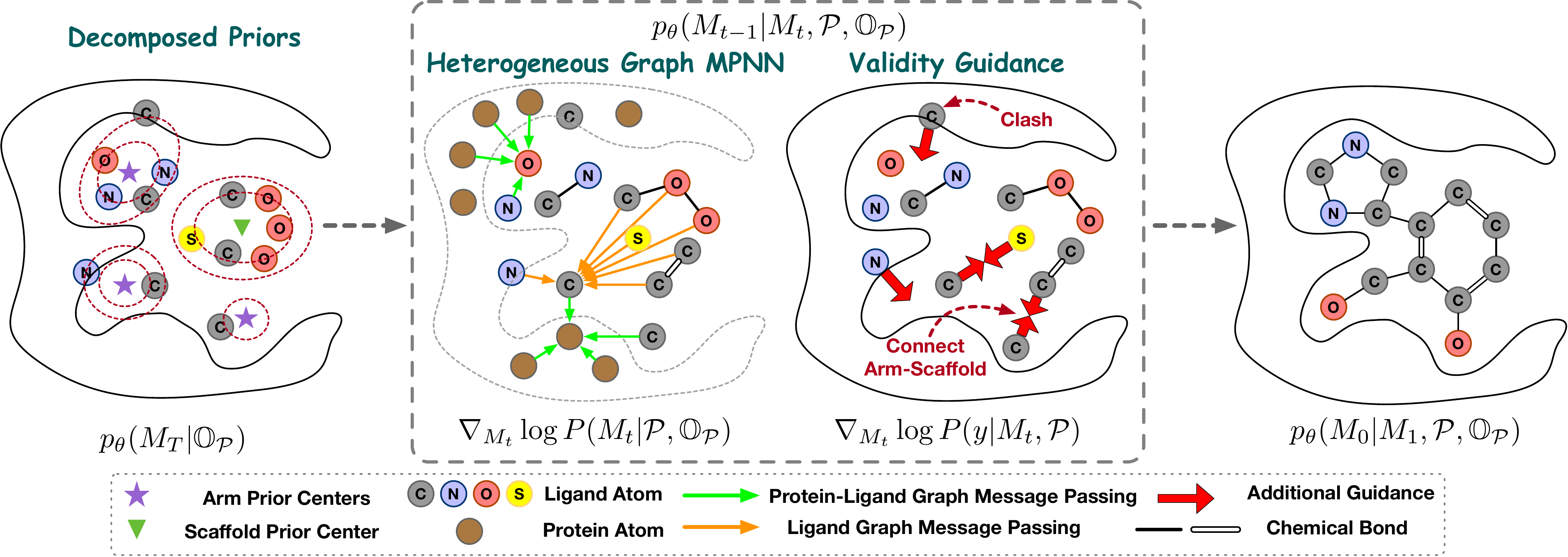}}
\vspace{-2mm}
\caption{Overview of the sampling process of DecompDiff. (a) The initial atoms are sampled from informative decomposed priors. (b) An equivariant network on heterogeneous graphs denoises atom coordinates, atom types and bond types simultaneously. (c) The validity guidance alleviates the protein-ligand clash problem and encourages arms and scaffold to connect. }
\label{fig:model}
\vspace{-8mm}
\end{center}
\end{figure*}

In this section, we present \method, which injects informative priors of the decomposed arms and scaffolds into a diffusion model for SBDD. We first define the SBDD task and introduce the standard diffusion model in Sec. \ref{sec:prelim}. Then, we show how to introduce decomposed priors into the diffusion model in Sec. \ref{sec:method_prior}. In Sec. \ref{sec:bond_diff}, we present bond diffusion and the network used for 3D molecular graph generation. Finally, in Sec. \ref{sec:guidance}, we describe an effective guided sampling approach to improve the validity and quality of the generated molecules.  

\subsection{Preliminaries}
\label{sec:prelim}
In SBDD, we are provided with a protein binding site, which can be represented as a set of $N_P$ atoms $\gP = \{ (\vx_{P}^{(i)}, \vv_{P}^{(i)}) \}_{i=1}^{N_P}$. The goal is to generate ligand molecules that can bind with the protein. Similarly, the ligand molecule can be represented as a set of $N_M$ atoms $\gM=\{(\vx_{M}^{(i)}, \vv_{M}^{(i)}) \}_{i=1}^{N_M}$. Here $\vx \in \R^3$ and $\vv \in \R^d$ denote the position and type of the atom respectively. The number of atoms $N_M$ can be sampled from an empirical distribution \citep{hoogeboom2022equivariant,guan20233d} or predicted by a neural network \citep{lin2022diffbp}, and is not involved in the diffusion process.
By denoting the ligand molecule as $M=[\rvx, \rvv]$ for brevity, where $\rvx\in \R^{N_M\times 3}$ and $\rvv\in \R^{N_M\times d}$, the SBDD task can be formulated as modeling the conditional distribution $p(M|\gP)$. 

Considering this task in the Denoising
Diffusion Probabilistic Model (DDPM) framework, a small
Gaussian noise is gradually injected into data as a Markov chain, leading to the following forward diffusion process:
\begin{equation}
\label{eq:method_standard_diffusion_process}
    q(M_{1:T}|M_0,\gP) = \prod_{t=1}^Tq(M_t|M_{t-1},\gP),
\end{equation}
where the data $M_0\sim p(M_0|\gP)$ and $M_1, M_2, \cdots, M_T$ is a sequence of latent variables induced by the diffusion process.
The reverse process, also known as the generative process, learns to recover data by iteratively denoising with a neural network parameterized by $\theta$ as follows: 
\begin{equation}
\label{eq:method_standard_reverse_process}
    p_{\theta}(M_{0:T-1}) = \prod_{t=1}^T p_{\theta}(M_{t-1}|M_t, \gP).
\end{equation}
The distribution of $p(M_T)$ induced by the forward diffusion process is Gaussian and works as the prior distribution during sampling. 

We can derive a tractable variational lower bound on the log-likelihood of data $M_0$, also known as evidence lower bound (ELBO), due to the property of standard Gaussian distribution. To align with the concept of loss, we can formulate it as follows:
\begin{equation}
\label{eq:method_standard_elbo}
    -\log p(M_0|\gP) \leq L_0 + \sum_{t=1}^{T-1}L_t + L_T,
\end{equation}
where $L_0 = -\mathbb{E}[\log p_\theta(M_0|M_1,\gP)]$ denotes the data likelihood, $L_T=D_{\text{KL}}(q(M_T|M_0,\gP)\Vert p_\theta(M_T))$ is the Kullback-Leibler (KL) divergence between the final distribution induced by the diffusion process $q(M_T|M_0,\gP)$ and the prior distribution $p_\theta(M_T)$, and $L_t = -D_{\text{KL}}(q(M_t|M_{t+1}, M_{0},\gP)\Vert p_\theta(M_t|M_{t+1},\gP))$. 
% Here, $D_{\text{KL}}$ denotes the Kullback-Leibler (KL) divergence.

\subsection{Diffusion Model with Decomposed Priors}
\label{sec:method_prior}
% Outline:
% * (Background)
% * Data-dependent prior formulation (forward, backward, ELBO)
% * How do we get the prior information
% * Decomposition-related guidance

Motivated by the natural decomposition of the ligand molecules, we decompose a ligand into fragments $\mathcal{K}$, consisting of arms $\mathcal{A}$ and scaffold $\mathcal{S}$ ($|\mathcal{A}|\geq1, |\mathcal{S}|\leq1, K=|\mathcal{K}|=|\mathcal{A}|+|\mathcal{S}|$).  From the perspective of machine learning, fragmentation provides natural clusters of ligand atoms. Intuitively, if we could leverage these natural clusters as informative prior, approximating $p_\theta$ would be easier than a non-informative Gaussian prior. Following the optimal prior hypothesis \citep{vignac2022digress}, we estimate the prior of each cluster with a Gaussian distribution, which is obtained by Maximum Likelihood Estimation (MLE) on the positions of its cluster members. 
Specifically, given a protein binding site, ligand atoms exhibit a multimodal distribution spatially. 
In the training phase, the prior can be obtained from reference ligands as described above. In the test phase where reference ligands are not always available or ideal, the prior can be obtained by human experts or rule-based algorithms. We provide more details for obtaining priors and fragmentation in \cref{sec:prior_generation} and \cref{sec:fragmentation}.

% Following the above description, we can obtain a set of data-dependent priors $\sO_\gP = \{\vmu^{\gP}_k, \mSigma^{\gP}_k\}_{k=1}^K$ and $\mH_\gP = [\eta^{\gP}_{ik}]_{N_M\times K}, \sum_{k=1}^K \eta_{ik}^\gP =1, \eta_{ik}^{\gP}\in\{0,1\} $ conditioned on the protein $\gP$. For brevity, we ignore the condition superscript $\gP$ in the following derivation. $\vmu_k \in \R^{3}$ is the prior center, $\mSigma_k \in \R^{3 \times 3}$ is the prior covariance matrix and $\eta_{ik} = 1$ indicates that the $i$th molecule atom corresponds to the $k$th prior. Note that an atom can only be assigned with one single prior. 

Following the above description, we can obtain a set of data-dependent priors $\sO_\gP = \{\vmu^{\gP}_{1:K}, \mSigma^{\gP}_{1:K}, \mH_\gP\}$, where $\vmu_k \in \R^{3}$ is the prior center, $\mSigma_k \in \R^{3 \times 3}$ is the prior covariance matrix and $\mH_\gP = \{\mathbf{\eta}^{\gP} \in \{0, 1\}^{N_M\times K} | \sum_{k=1}^K \eta_{ik}^\gP =1\}$ is the prior-atom mapping conditioned on the protein $\gP$, i.e., $\eta_{ik} = 1$ indicates that the $i$-th molecule atom corresponds to the $k$-th prior. Note that an atom can only be assigned with one single prior. 
% For brevity, we ignore the condition superscript $\gP$ in the following derivation. 
Next, we will describe how the diffusion process, generative process and ELBO will be adjusted accordingly when including the decomposed priors.

Our goal is to model the conditional molecular distribution $p_\theta(M | \gP) = p_\theta(M | \gP,\sO_\gP)$. 
Since our informative prior is defined in the 3D coordinate space, we only consider the diffusion and generative process of ligand atom positions $\rvx$ and omit the ligand atom types $\rvv$ for clarity in the following derivation. 
The condition $\gP$ in $\sO_\gP$ is also omitted.
% We denote $\{\rvx_t^{(i)}\}_{i=1}^{N_M}$, i.e. the positions of all atoms in a molecule at time step $t$, as $\rvx_t$. 

To distinguish the difference induced by the decomposed priors from the standard ones, we highlight the critical differences of equations in \blue{blue}. To achieve SE(3)-equivariance, we apply a center shifting operation \citep{xu2022geodiff} on every atom according to its corresponding prior center, denoted as $\Tilde{\rvx}^{(i)}_{t,{\blue{k}}}=\rvx^{(i)}_{t}-\vmu_{\blue{k}}$. We follow the definitions and notations related to the noise schedule $\alpha_t, \beta_t, \Bar{\alpha}_t,\Tilde{\beta}_t$ in \citet{ho2020denoising}. With data-dependent prior as the additional context, the diffusion and generative process in \cref{eq:method_standard_diffusion_process,eq:method_standard_reverse_process} can be extended as follows:
\begin{small}
\begin{equation}
\label{eq:forward_decomp}
q(\rvx_t | \rvx_{t-1}, \gP)
= \prod_{i=1}^{N_M}\blue{\!\sum_{k=1}^{K} \eta_{ik}} \gN(\Tilde{\rvx}_{t,{\blue{k}}}^{(i)} ;\Tilde{\rvx}_{t-1,{\blue{k}}}^{(i)}, \beta_t \blue{\mSigma_k})
\end{equation}
\begin{equation}
\label{eq:reversed_decomp}
p_\theta(\rvx_{t-1} \vert \rvx_t, \rvx_0,\gP)
= \prod_{i=1}^{N_M}\blue{\sum_{k=1}^{K}\eta_{ik}}
    \gN(\Tilde{\rvx}_{t-1,{\blue{k}}}^{(i)};
    \Tilde{\vmu}_t(\Tilde{\rvx}_{t,{\blue{k}}}^{(i)}, \Tilde{\rvx}_{0,{\blue{k}}}^{(i)}), \Tilde{\beta}_t \blue{\mSigma_k})
\end{equation}
\end{small}
where $\Tilde{\vmu}_t(
\Tilde{\rvx}_{t,{\blue{k}}}^{(i)}
,
\Tilde{\rvx}_{0,{\blue{k}}}^{(i)})
=
\frac{\sqrt{\alpha_t}(1-\Bar{\alpha}_{t-1})}{1-\Bar{\alpha}_t}
\Tilde{\rvx}_{t,{\blue{k}}}^{(i)} 
+
\frac{\sqrt{\Bar{\alpha}_{t-1}}\beta_t}{1-\Bar{\alpha}_t}
\Tilde{\rvx}_{0,{\blue{k}}}^{(i)} 
$.

The prior distribution induced by the forward diffusion process can be derived as follows:
\begin{small}
\begin{equation}
\label{eq:prior_decomp}
p(\rvx_T \vert\gP )
=
\prod_{i=1}^{N_M} \blue{\sum_{k=1}^K \eta_{ik} }\gN(\rvx_{T}^{(i)};\vmu_{\blue{k}}, \blue{\mSigma_k}).
\end{equation}
\end{small}

% The reversed diffusion process can be extended as follows:
% \begin{equation}
% \begin{split}
% \label{eq:reversed_decomp}
%     & q(\rvx_{t-1} \vert
%     \rvx_t,
%     \rvx_0,\gP)
%     \\ 
%     & = \prod_{i=1}^{N_M}\blue{\sum_{k=1}^{K}\eta_{ik}}
%     \gN(\Tilde{\rvx}_{t-1,{\blue{k}}}^{(i)};
%     \Tilde{\vmu}_t(\Tilde{\rvx}_{t,{\blue{k}}}^{(i)}, \Tilde{\rvx}_{0,{\blue{k}}}^{(i)}), \Tilde{\beta}_t \blue{\mSigma_k})
% \end{split}
% \end{equation}

The evidence lower bound (ELBO) is still traceable as in \cref{eq:method_standard_elbo} where $L_0$, $L_t$, and $L_T$ are as follows: 
\begin{small}
\begin{equation}
\label{eq:pos_l0}
\begin{aligned}
L_0  = 
-\mathbb{E}_{\rvx_0,\rvx_1}
\Bigg[  \sum_{i=1}^{N_M}\blue{\sum_{k=1}^{K}\eta_{ik}}
\bigg[
    &\frac{1}{2\Tilde{\beta}_0}
    \big\Vert 
    {\rvx}_{0}^{(i)}
    -
    \hat{\rvx}_{0,1}^{(i)}
    \big\Vert^2_{\blue{\mSigma_k^{-1}}} \\
 &+\frac{1}{2}\ln{\det\left({\blue{\mSigma_k^{-1}}}\right)}
    \bigg] 
    \Bigg] + C_0,
% \end{aligned}
% \end{split}
\end{aligned}
\end{equation}
\begin{equation}
\label{eq:pos_lt}
% \begin{split}
L_t = 
% D_{\text{KL}}\left(q(\rvx_{t} \vert \rvx_{t+1}, \rvx_{0}, \gP)
% \Vert
% p_{\theta}(\rvx_{t} \vert \rvx_{t+1}, \gP)\right)
% \\
\mathbb{E}_{\rvx_0,\rvx_t}
\left[ 
\gamma_t
\sum_{i=1}^{N_M}\blue{\sum_{k=1}^{K}\eta_{ik}}
\left\Vert 
{\rvx}_{0}^{(i)}
-
\hat{\rvx}_{0,t+1}^{(i)}
\right\Vert^2_{\blue{\mSigma_k^{-1}}}
\right] + C_t,
% \end{split}
\end{equation}
\begin{equation}
\label{eq:pos_lT}
L_T = \mathbb{E}_{\rvx_0}
\left[ 
\frac{\Bar{\alpha}_T}{2}
\sum_{i=1}^{N_M}\blue{\sum_{k=1}^{K}\eta_{ik}}
\left\Vert 
\Tilde{\rvx}_{0,{\blue{k}}}^{(i)}
\right\Vert^2_{\blue{\mSigma_k^{-1}}}
\right] + C_T,
\end{equation}
\end{small}
where $C_0$, $C_t$, and $C_T$ are constant terms, and $\gamma_t = \frac{\Bar{\alpha}_{t-1}\beta_t^2}{2(1-\Bar{\alpha}_t)^2\Tilde{\beta}_t}$. $\hat{\rvx}_{0,t}^{(i)}=f_\vtheta
(\rvx_{t}^{(i)}, t, \gP)$ is implemented by a SE(3)-equivariant neural network parameterized by $\vtheta$. Note that the input of the neural network $\rvx_t^{(i)} = \sqrt{\Bar{\alpha}_t} \blue{\sum_{k=1}^K}\Tilde{\rvx}_{0,\blue{k}}^{(i)} + \sqrt{1-\Bar{\alpha}_t} \rvepsilon_{\blue{k}} + \vmu_{\blue{k}}$ is different from that under the standard prior.
% due to the different diffusion process defined in \cref{eq:forward_decomp}.

% We analyze the ELBO based on the decomposed priors theoretically in \cref{sec:proof} and provide insights for the reason why such priors could induce tighter ELBO, though the theoretical analysis is under some restrictive assumptions.
\begin{proposition}\label{prop:elbo}
Let $-\text{\rm ELBO}_{\text{\rm decomp}}(\vtheta)$ and $-\text{\rm  ELBO}_{\text{\rm standard}}(\vtheta)$ denote the $-\text{\rm  ELBO}$ losses under the decomposed prior and the standard Gaussian prior respectively. 
Suppose that $f_\vtheta$ is a simple graph neural network with an equivariant linear layer. If the decomposed prior aligns with data distribution, 
we have $\min_{\vtheta}-\text{\rm ELBO}_{\text{\rm decomp}}(\vtheta)\leq \min_{\vtheta} -\text{\rm ELBO}_{\text{\rm standard}}(\vtheta)$.
% the minimum value of $-\text{ELBO}_{\text{decomp}}(\vtheta)$ under the decomposed prior is always less than the minimum value of $-\text{ELBO}_{\text{standard}}(\vtheta)$ under the standard Gaussian prior.
\end{proposition}

\cref{prop:elbo} provides insights into why our priors can induce better results. Please see \cref{sec:proof} for its proof.

\subsection{Bond Diffusion and Model Architecture}\label{sec:bond_diff}

\paragraph{Introducing Bond Diffusion} Existing diffusion models for 3D molecule generation \cite{hoogeboom2022equivariant, guan20233d, schneuing2022structure, lin2022diffbp} only consider generating atom coordinates and atom types with neural networks, while adding bonds by a post-processing algorithm, such as that implemented in OpenBabel \citep{o2011open}. Ideally, this paradigm can work well if the atom coordinates are predicted accurately. However, since the distributions of bond distances are very sharp and a small error may lead to totally different molecules, adding bonds based on imperfect atom coordinates with a post-processing algorithm is not always reliable. 

To address this problem, we develop a new diffusion framework for 3D molecular \textit{graph} generation which also considers bonds in the dynamics. Specifically, we extend the molecular representation as $\gM=\{(\vx_i, \vv_i, \vb_{ij})\}_{i, j \in \{1, \dots,N_M\}}$. 
We apply the discrete diffusion \citep{hoogeboom2021argmax} for bond types, similar to how we model the atom types. The forward diffusion becomes as follows:
\begin{small}
\begin{equation}
\begin{aligned}
% \label{eq:detail_diff}
& q(M_t | M_{t-1}, \gP) = \gN(\rvx_t; \sqrt{1-\beta_t} \rvx_{t-1}, \beta_t \rmI) \\ & \cdot \gC(\rvv_t | (1-\beta_t) \rvv_{t-1} + \beta_t / K_a) \cdot \gC(\rvb_t | (1-\beta_t) \rvb_{t-1} + \beta_t / K_b),
\end{aligned}
\end{equation}
\end{small}
where $K_a$ and $K_b$ are the number of atom types and bond types respectively. Benefiting from decomposing the drug space, we do not have to build the bond dynamics on the fully connected graph of the ligand molecule, but instead inside of arms/scaffold and between arms and scaffold: $\rvb = \{b_{ij} | i \in \mathcal{K}_n, j \in \mathcal{K}_n \}_{n=1:|\mathcal{K}|} \cup \{b_{ij} | i \in \mathcal{A}_n, j \in \mathcal{S} \}_{n=1:|\mathcal{A}|}$
% Note that although we decompose the molecular space, the bond dynamics is also built on the fully connected graph of the whole molecule to model the connections between arms and scaffold. 

\paragraph{Equivariant Network with Nodes and Edges Update} 
Inspired by recent progress in equivariant neural networks~\citep{thomas2018tensor, fuchs2020se, satorras2021n, guan2021energy}, we propose a new equivariant neural network to denoise 3D molecular graph. Specifically, we maintain both node-level and edge-level hidden states in the neural network to better reflect the bond diffusion, unlike the commonly used EGNN \citep{satorras2021n} where only node-level representation is considered.

We first build a $k$-nearest neighbors (knn) graph $\mathcal{G}_{K}$ upon ligand atoms and protein atoms to model the protein-ligand interaction: 
% Denoting atoms' hidden states as $\rvh$, we perform message passing on this graph as follows: 
{\small
\begin{equation}
\label{eq:pl_message}
\begin{aligned}
    \Delta\rvh_{K, i}  &\leftarrow \sum_{j\in \mathcal{N}_K(i)} \phi_{m_K} (\rvh_i, \rvh_j, \|\rvx_i - \rvx_j\|, E_{ij}, t), 
\end{aligned}
\end{equation}
\normalsize}
where $\rvh$ is the atom's hidden state, $\mathcal{N}_K(i)$ is the neighbors of $i$ in $\mathcal{G}_K$, $E_{ij}$ indicates the edge $ij$ is a protein-protein, ligand-ligand or protein-ligand edge.

We also build a fully/partially connected ligand graph $\mathcal{G}_{L}$ upon ligand atoms to model the interaction inside the ligand:
{\small
\begin{equation}\label{eq:ll_message}
\begin{aligned}
    \rvm_{ij}  &\leftarrow \phi_d(\|\rvx_i - \rvx_j\|, \rve_{ij}) \\ 
    \Delta\rvh_{L, i}  &\leftarrow \sum_{j\in{\mathcal{N}_L(i)}} \phi_{m_L} (\rvh_i, \rvh_j, \rvm_{ji}, t),
\end{aligned}
\end{equation}
\normalsize}
% {\small
% \begin{align}\label{eq:ll_message}
%     \hat{\rve}_{ij}  &\leftarrow \phi_d(\|\rvx_i - \rvx_j\|, \rve_{ij}) \\ 
%     \rvm_{L, i}  &\leftarrow \sum_{j\in{\mathcal{V}_{L}}, i\neq j} \phi_{m_L} (\rvh_i, \rvh_j, \hat{\rve}_{ji}, t),
% \end{align}
% \normalsize}
where $\rve$ is the bond's hidden state. 
Based on the messages aggregated from heterogeneous graphs, we update the atom's hidden state as \cref{eq:node_update}:
{\small
\begin{equation}
\label{eq:node_update}
\begin{aligned}
    \rvh_i  \leftarrow \rvh_i + \phi_h(\Delta\rvh_{K, i} + \Delta\rvh_{L, i}).
\end{aligned}
\end{equation}
\normalsize}

We update the bond's hidden state following a directional message passing \citep{yang2019analyzing, gasteiger2020directional} schema as \cref{eq:edge_update}:
{\small
\begin{equation}
\label{eq:edge_update}
\begin{aligned}
    \rve_{ji}  \leftarrow \sum_{k\in \mathcal{N}_L(j) \backslash \{i\}} \phi_e(\rvh_i, \rvh_j, \rvh_k, \rvm_{kj}, \rvm_{ji}, t).
\end{aligned}
\end{equation}
\normalsize}

Finally, the atom positions of ligand molecules are updated as \cref{eq:x_update}:
{\small
\begin{equation}
\label{eq:x_update}
\begin{aligned}
    \Delta\rvx_{K, i}  &\leftarrow \sum_{j\in{\mathcal{N}_K(i)}} (\rvx_j-\rvx_i) \phi_{x_K} (\rvh_i, \rvh_j, \|\rvx_i - \rvx_j\|, t) \\
    \Delta\rvx_{L, i}  &\leftarrow \sum_{j\in{\mathcal{N}_L(i)}} (\rvx_j-\rvx_i) \phi_{x_L}(\rvh_i, \rvh_j, \|\rvx_j-\rvx_i\|, \rvm_{ji}, t) \\
    \rvx_i  &\leftarrow \rvx_i + (\Delta\rvx_{K, i} + \Delta\rvx_{L, i}) \cdot \mathds{1}_{\text{mol}},
\end{aligned}
\end{equation}
\normalsize}
where $\mathds{1}_{\text{mol}}$ is the indicator of ligand atoms since we assume the protein atoms are fixed as the context. 

We obtain the initial atom hidden state $\rvh^0$ and bond hidden state $\rve^0$ by two embedding layers that encode atom, bond and decomposition information. The final hidden states $\rvh^L$ and $\rve^L$ are fed into two MLPs to obtain the predicted atom type $\hat{\rvv}_i=\text{softmax}(\text{MLP}(\rvh^L_i))$ and bond type $\hat{\rvb}_{ij}=\text{softmax}(\text{MLP}(\rve^L_{ij} + \rve^L_{ji}))$. Since atom type and bond type are subject to categorical distributions, we can directly compute the KL divergence between their estimated posteriors and the truth posteriors as losses:
\begin{small}
\begin{equation}
\label{eq:atom_type_loss}
    L_{t}^{(v)} = \sum_{k=1}^{K_a} \bm{c}(\rvv_t, \rvv_0)_k \log \frac{\bm{c}(\rvv_t, \rvv_0)_k}{\bm{c}(\rvv_t, \hat\rvv_0)_k}
    ,
\end{equation}
\begin{equation}
\label{eq:bond_type_loss}
    L_{t}^{(b)} = \sum_{k=1}^{K_b} \bm{c}(\rvb_t, \rvb_0)_k \log \frac{\bm{c}(\rvb_t, \rvb_0)_k}{\bm{c}(\rvb_t, \hat\rvb_0)_k}
    ,
\end{equation}
\end{small}
where ${\bm c}(\rvv_t, \rvv_0) = \bm{c}^\star / \sum_{k=1}^{K_a} c_k^\star$ and $\bm{c}^\star(\rvv_t, \rvv_0) = [\alpha_t\rvv_t + (1 - \alpha_t) / K_a] \odot [\bar\alpha_{t-1}\rvv_0 + (1 - \bar\alpha_{t-1}) / K_a]$. ${\bm c}(\rvb_t, \rvb_0)$ is defined in the similar way. 
Combined with the decomposed atom position loss described in Sec. \ref{sec:method_prior}, the final loss is a weighted sum of atom position loss, atom type loss and bond type loss: $L=L_t^{(x)} + \gamma_a L_t^{(v)} + \gamma_b L_t^{(b)}$, where $L_t^{(x)}$ is from \cref{eq:pos_l0,eq:pos_lt,eq:pos_lT}. More implementation details can be found in Appendix \ref{sec:app:model_details}.

\subsection{Validity Guidance}
\label{sec:guidance}

To further improve the validity and quality of generated molecules, we introduce additional drift terms to guide the Langevin dynamics during sampling. Inspired by classifier guidance for conditional sampling of DDPM \citep{dhariwal2021diffusion}, we design several guidance terms $\nabla_{\vx_t} \log{P(y \vert \vx_t)}$ which are applied during the sampling phase to constrain the generated molecules in a specific domain $y$ as follows:
\begin{equation}
    \nabla_{\vx_t} \log{P(\vx_t \vert y)} =  \nabla_{\vx_t} \log{P(\vx_t)} + \nabla_{\vx_t} \log{P(y \vert \vx_t)}.
\end{equation}

We consider the validity of generated molecules from two aspects: (a) the arms and scaffold should be connected to form a complete molecule; (b) there should not be a clash between the generated molecule and protein surface. We will describe the guidance design from these two aspects detailedly in the following and leave related derivation in \cref{sec:guidance_derivation}. 

We assert the existence of connections between all arms and scaffold if the following inequality holds for $n=1:|\mathcal{A}|$,
\begin{equation}
    \rho_{\min} \leq \min_{i\in \mathcal{A}_n, j\in \mathcal{S}} \Vert \vx^{(i)} - \vx^{(j)} \Vert_2 \leq \rho_{\max},
\end{equation}
where $\rho_{\min}$ and $\rho_{\max}$ are hyperparameters approximately representing the range of a bond length and set to 1.2{\AA } and 1.9{\AA } respectively in practice. The arms-scaffold drift can be derived as follows:
\begin{small}
    \begin{equation}
    \label{eq:arms_scaffold_drift_final}
     -\nabla_{\vx_t}\sum_{n=1}^{|\mathcal{A}|}[
        \xi_2 \max{(0,d_t^{(n)}-\rho_{\max})}
        +
        \xi_1 \max{(0,\rho_{\min}-d_t^{(n)})}
        ],
\end{equation}
\end{small}
where $d_t^{(n)}=\min_{i\in \mathcal{A}_n, j\in \mathcal{S}} \Vert \vx_t^{(i)} - \vx_t^{(j)} \Vert_2$ and $\xi_1,\xi_2>0$ are constant coefficients that control the strength of drift.
 
% A non-intersection loss is introduced as a regularization term in docking methods based on deep learning \citep{ganea2021independent,stark2022equibind} to avoid such unrealistic cases. This loss is then introduced to structure-based 3D molecule generation \citep{lin2022diffbp} to consider the inductive bias of non-intersection between the proteins and reconstructed binding molecules during training.

% However, introducing an additional loss as a regularization term during training requires a lot of extra effort in adjusting hyperparameters to balance the loss and the others \orange{(yy: what does "the others" stand for?)}. A more efficient way is to only adjust the Langevin dynamics during sampling by introducing an additional drift term. 

Too short distances or clashes between the atoms of generated molecules and proteins will induce abnormal Van der Waals forces.
Following \citet{sverrisson2021fast,ganea2021independent}, we choose $\{\vx\in\R^3:S(\vx)=\gamma\}$ where $S(\vx)=-\sigma \ln{(\sum_{j=1}^{N_P} \exp{(-\Vert \vx-\vx_P^{(j)} \Vert^2/\sigma)} )}$ as the descriptor of the protein surface. Recall that $\{\vx_P^{(j)}\}_{j=1}^{N_P}$ represents the set of positions of protein atoms. The clash drift can be derived as follows:
\begin{equation}
\label{eq:clash_drift_final}
    -\nabla_{\vx_t} \xi_3 \sum_{i}^{N_M} \max{(0, \gamma-S(\vx_t^{(i)}))}, 
\end{equation}
where $\xi_3>0$ is the constant coefficient that controls the strength of drift.

\section{Experiments}

\subsection{Experimental Setup}
\paragraph{Dataset} Following the previous work \cite{luo20213d, peng2022pocket2mol}, we trained our model on the CrossDocked2020 dataset \citep{francoeur2020three}. We use the same dataset preprocessing and splitting procedure as \citet{luo20213d}, where the $22.5$ million docked binding complexes are first refined to only keep high-quality docking poses (RMSD between the docked pose and the ground truth $<1$\AA) and diverse proteins (sequence identity $<$ $30\%$), and then $100,000$ complexes are selected for training and $100$ novel proteins are selected as references for testing.

\paragraph{Baselines}
We compare our model with various representative baselines: \textbf{liGAN} \citep{ragoza2022chemsci} is a conditional VAE model which uses 3D CNN to encode and generate voxelized atomic density. \textbf{AR} \citep{luo20213d}, \textbf{Pocket2Mol} \citep{peng2022pocket2mol} and \textbf{GraphBP} \citep{liu2022graphbp} are GNN-based methods that generate 3D molecules atom by atom in an autoregressive manner. \textbf{TargetDiff} \citep{guan20233d} is a diffusion-based method which generates atom coordinates and atom types in a non-autoregressive way, but the prior distribution is a standard Gaussian and bonds are generated with a post-processing algorithm. 

\begin{figure}[t]
\begin{center}
\centerline{\includegraphics[width=\linewidth]{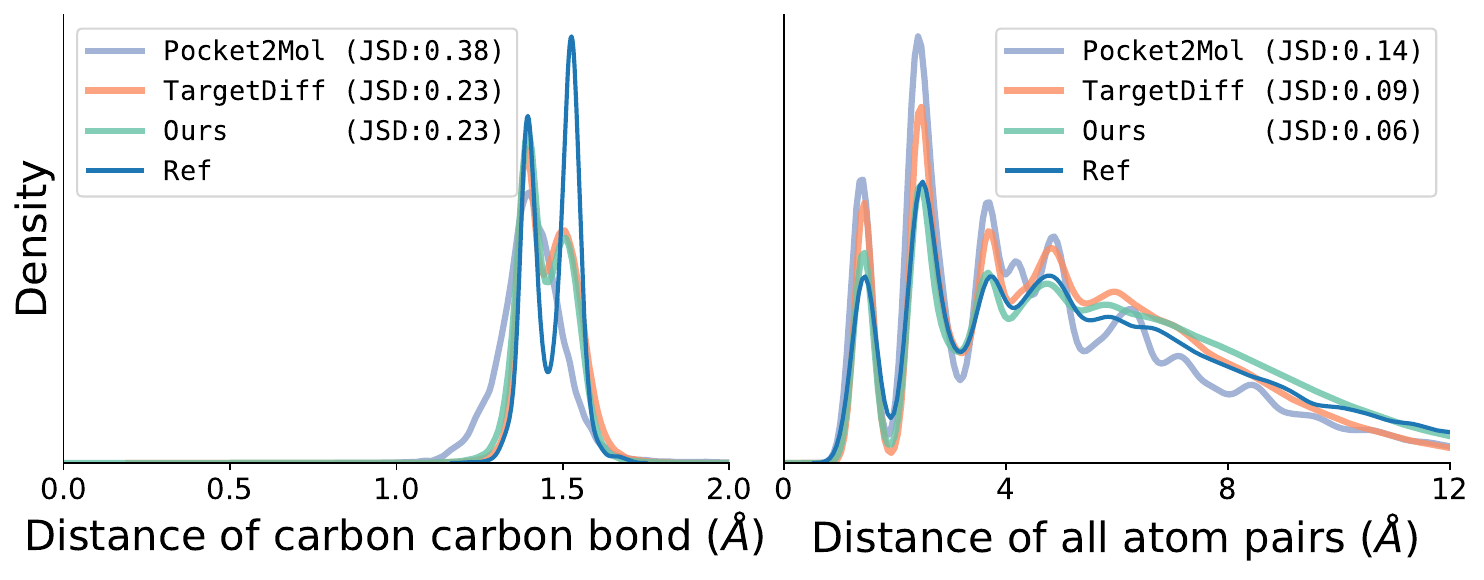}}
\vspace{-3mm}
\caption{Comparing the distribution for distances of carbon-carbon pairs (left) and all-atom (right) for reference molecules in the test set and model-generated molecules. Jensen-Shannon divergence (JSD) between two distributions is reported. }
\label{fig:atom_jsd}
\vspace{-5mm}
\end{center}
\end{figure}

\begin{table}[t]
    \centering
    \caption{Jensen-Shannon divergence between bond distance distributions of the reference molecules and the generated molecules, and lower values indicate better performances. ``-'', ``='', and ``:'' represent single, double, and aromatic bonds, respectively. We highlight the best two results with \textbf{bold text} and \underline{underlined text}, respectively.}
    % \vspace{2mm}
    \small
    \begin{adjustbox}{width=0.5\textwidth}
\begin{tabular}{ccccccc}
\toprule
{Bond} & liGAN & GraphBP & AR & \thead{Pocket2 \\ Mol} & \thead{Target \\ Diff} & Ours \\
\midrule
C$-$C & 0.601 & 0.368 & 0.609 & 0.496 & \underline{0.369} & \textbf{0.359} \\
C$=$C & 0.665 & 0.530 & 0.620 & 0.561 & \textbf{0.505} & \underline{0.537} \\
C$-$N & 0.634 & 0.456 & 0.474 & 0.416 & \underline{0.363} & \textbf{0.344} \\
C$=$N & 0.749 & 0.693 & 0.635 & 0.629 & \textbf{0.550} & \underline{0.584} \\
C$-$O & 0.656 & 0.467 & 0.492 & 0.454 & \underline{0.421} & \textbf{0.376} \\
C$=$O & 0.661 & 0.471 & 0.558 & 0.516 & \underline{0.461} & \textbf{0.374} \\
C$:$C & 0.497 & 0.407 & 0.451 & 0.416 & \underline{0.263} & \textbf{0.251} \\
C$:$N & 0.638 & 0.689 & 0.552 & 0.487 & \textbf{0.235} & \underline{0.269} \\
\bottomrule
\end{tabular}
\label{tab:bond_jsd}
    \end{adjustbox}
    \vspace{-4mm}
\end{table}
% \todoq{need to squeeze the white space after Table 1.}

\begin{table}[t]
    \centering
    \caption{Jensen-Shannon divergence between bond distance distributions of the reference molecules and the generated molecules, and lower values indicate better performances. We highlight the best two results with \textbf{bold text} and \underline{underlined text}, respectively.}
    % \vspace{2mm}
    \small
    \begin{adjustbox}{width=0.5\textwidth}
\begin{tabular}{ccccccc}
\toprule
{Angle} & liGAN & GraphBP & AR & \thead{Pocket2 \\ Mol} & \thead{Target \\ Diff} & Ours \\
\midrule
CCC & 0.598 & 0.424 & 0.340 & \underline{0.323} & 0.328 & \textbf{0.314}  \\
CCO & 0.637 & \underline{0.354} & 0.442 & 0.401 & 0.385 & \textbf{0.324}  \\ 
CNC & 0.604 & 0.469 & 0.419 & \textbf{0.237} & 0.367 & \underline{0.297}  \\
OPO & 0.512	& 0.684 & 0.367	& \underline{0.274}	& 0.303	& \textbf{0.217}  \\
NCC & 0.621	& 0.372	& 0.392	& \underline{0.351}	& 0.354	& \textbf{0.294}  \\
CC=O & 0.636 & 0.377 & 0.476 & \underline{0.353} & 0.356 & \textbf{0.259}  \\
COC  & 0.606 & 0.482 & 0.459 & \textbf{0.317} & 0.389 & \underline{0.339} \\
\bottomrule
\end{tabular}
\label{tab:bond_angle_jsd} 
    \end{adjustbox}
    \vspace{-4mm}
\end{table}

\begin{table*}[t]
    \vspace{-4mm}
    \centering
    \caption{Summary of different properties of reference molecules and molecules generated by our model and other baselines. ($\uparrow$) / ($\downarrow$) denotes a larger / smaller number is better. Top 2 results are highlighted with \textbf{bold text} and \underline{underlined text}, respectively. 
    % For liGAN and GraphBP, AutoDock Vina could not parse some generated atom types and thus we use QVina \citep{alhossary2015fast} to perform docking. 
    }
    \begin{adjustbox}{width=1\textwidth}
    \renewcommand{\arraystretch}{1.2}
\begin{tabular}{l|cc|cc|cc|cc|cc|cc|cc|c}
\toprule
% \diagbox{Model}{Metric} 
\multirow{2}{*}{Methods} & \multicolumn{2}{c|}{Vina Score ($\downarrow$)} & \multicolumn{2}{c|}{Vina Min ($\downarrow$)} & \multicolumn{2}{c|}{Vina Dock ($\downarrow$)} & \multicolumn{2}{c|}{High Affinity ($\uparrow$)} & \multicolumn{2}{c|}{QED ($\uparrow$)}   & \multicolumn{2}{c|}{SA ($\uparrow$)} & \multicolumn{2}{c|}{Diversity ($\uparrow$)} & Success Rate ($\uparrow$) \\
 & Avg. & Med. & Avg. & Med. & Avg. & Med. & Avg. & Med. & Avg. & Med. & Avg. & Med. & Avg. & Med. & Avg. \\
\midrule
Reference   & -6.36 & -6.46 & -6.71 & -6.49 & -7.45 & -7.26 & -  & - & 0.48 & 0.47 & 0.73 & 0.74 & - & - & 25.0\%   \\
\midrule
liGAN       & - & - & - & - & -6.33 & -6.20 & 21.1\% & 11.1\% & 0.39 & 0.39 & 0.59 & 0.57 & 0.66 & 0.67 & 3.9\% \\

GraphBP     & - & - & - & - & -4.80 & -4.70 & 14.2\% & 6.7\% & 0.43 & 0.45 & 0.49 & 0.48 & \textbf{0.79} & \textbf{0.78} &  0.1\%  \\

AR          & \textbf{-5.75} & -5.64 & -6.18 & -5.88 & -6.75 & -6.62 & 37.9\% & 31.0\% & \underline{0.51} & \underline{0.50} & \underline{0.63} & \underline{0.63} & 0.70 & 0.70 & 7.1\% \\

Pocket2Mol  & -5.14 & -4.70 & -6.42 & -5.82 & -7.15 & -6.79 & 48.4\% & 51.0\% & \textbf{0.56} & \textbf{0.57} & \textbf{0.74} & \textbf{0.75} & 0.69 & 0.71 & \underline{24.4\%} \\

TargetDiff  & -5.47 & \textbf{-6.30} & \underline{-6.64} & \underline{-6.83} & \underline{-7.80} & \underline{-7.91} & \underline{58.1\%} & \underline{59.1\%} & 0.48 & 0.48 & 0.58 & 0.58 & \underline{0.72} & \underline{0.71} & 10.5\% \\

\method  & \underline{-5.67} & \underline{-6.04} & \textbf{-7.04} & \textbf{-7.09} & \textbf{-8.39} & \textbf{-8.43} & \textbf{64.4\%} & \textbf{71.0\%} & 0.45 & 0.43 & 0.61 & 0.60 & 0.68 & 0.68 & \textbf{24.5\%} \\
\bottomrule
\end{tabular}
\renewcommand{\arraystretch}{1}

    \end{adjustbox}\label{tab:mol_prop}
\end{table*}

\paragraph{Evaluation}
We evaluate the generated molecules from two perspectives: \textbf{molecular conformation} and \textbf{target binding affinity and molecular properties}. 
% The premise for SBDD is that the model could generate accurate ligand conformations to demonstrate an understanding of molecular validity and stability.
% we first plot the carbon-carbon bond distance distribution and all-atom pairwise distance distribution of the generated molecules.
% in Figure \ref{fig:atom_jsd}. 
In terms of molecular conformation, we compute the Jensen-Shannon divergences (JSD) in atom/bond distance distributions between the reference molecules and the generated molecules. 
We employ AutoDock Vina \citep{eberhardt2021autodock} to estimate the target binding affinity, following the same setup as \citet{luo20213d, ragoza2022chemsci}. 
We collect all generated molecules across 100 test proteins and report the mean and median of affinity-related metrics~(\textit{Vina Score}, \textit{Vina Min}, \textit{Vina Dock}, and \textit{High Affinity}) and property-related metrics~(drug-likeness \textit{QED} \citep{bickerton2012quantifying}, synthesizability \textit{SA} \citep{ertl2009estimation}, and \textit{diversity}). 
Vina Score directly estimates the binding affinity based on the generated 3D molecules, Vina Min conducts a local structure minimization before estimation, Vina Dock involves a re-docking process and reflects the best possible binding affinity, and High Affinity measures the percentage of how many generated molecules binds better than the reference molecule per test protein.
% Therefore, \textit{Vina Score} and \textit{Vina Min} are more direct metrics for evaluating the end-to-end 3D generation performance
Following \citet{yang2021knowledge, long2022zero}, we further report the percentage of molecules which pass certain criteria (QED $> 0.25$, SA $> 0.59$, Vina Dock $< -8.18$) as \textit{Success Rate} to comprehensively evaluate the target binding affinity and the molecular properties, given the fact that practical drug design also requires the generated molecules to be drug-like, synthesizable, and maintain high binding affinity simultaneously \citep{jin2020multi, xie2021mars}.
The thresholds used for QED and SA are computed as the 10th percentile of molecules in the DrugCentral database \cite{ursu2016drugcentral}, which are all pharmaceutical or under clinical trials.

\subsection{Main Results}
% \paragraph{Molecular Conformation.} 
First, we compare our model and the representative methods in terms of molecular conformation. 
We compute different bond distance and bond angle distributions of the generated molecules and compare them against the corresponding reference empirical distributions in \cref{tab:bond_jsd,tab:bond_angle_jsd}, and plot the carbon-carbon bond distance distribution and all-atom pairwise distance distribution of the generated molecules in \cref{fig:atom_jsd}. 
We see in \cref{fig:atom_jsd} that \method achieves the lowest JSD of $0.23$ and $0.06$ to reference in the carbon-carbon bond distance distribution and all-atom pairwise distance distribution of the generated molecules, indicating it captures real atomic distances well. Such performance is better than Pocket2Mol and TargetDiff, two strong baselines, while Pocket2Mol and TargetDiff have a biased estimation for carbon-carbon bond distribution and the long-range (especially $10-12$ \AA) atomic distance.
A similar observation can also be found in \cref{tab:bond_jsd,tab:bond_angle_jsd}.  
Our model has a comparable performance with TargetDiff and is better than all other baselines by a clear margin, showing the strong potential of our proposed method for generating stable molecular conformations directly.

Then we evaluate the effectiveness of our model in terms of binding affinity and molecular properties. 
We can see in \cref{tab:mol_prop} that our \method outperforms baselines by a large margin in affinity-related metrics. 
Specifically, \method surpasses the strong baseline TargetDiff by $6\%$ and $12\%$ in Avg. and Med. High Affinity, and around $0.40$ in Vina Min and Vina Dock.
All these gains clearly indicate that our decomposed priors are helpful for improving the potential of diffusion models for generating molecules with better target binding affinity.

We also see there is a trade-off between the property-related metrics~(QED, SA) and affinity-related metrics.
Like TargetDiff, our model falls behind the SOTA autoregressive model Pocket2Mol in QED and SA scores. 
Nevertheless, it is worth mentioning that such property-related metrics are often applied as rough screening metrics in real drug discovery scenarios as long as they fall into a reasonable range. 
Instead of directly comparing them in numerical value, a recommended evaluation strategy is using the Success Rate~\citep{jin2020multi, xie2021mars} to reflect the molecular properties comprehensively. 
In such a context, our \method achieves a $24.5\%$ Success Rate, which is comparable to Pocket2Mol and clearly outperforms TargetDiff. We also show some visualization results in Appendix \ref{sec:examples_appendix} to compare TargetDiff and \method.

\begin{figure*}[t]
% \vskip 0.2in
\vspace{-1mm}
\begin{center}
\centerline{\includegraphics[width=0.98\textwidth]{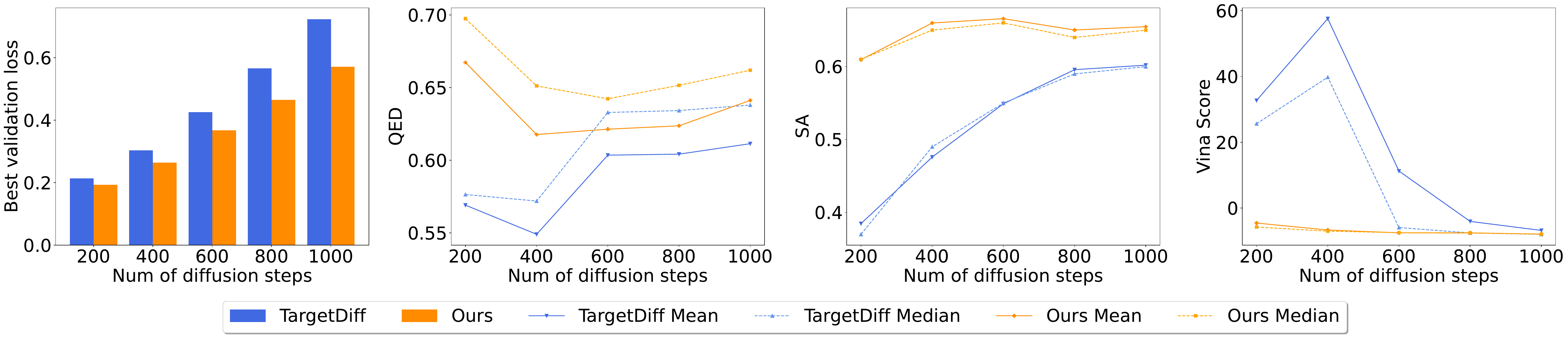}}
        %     \subfigure{
        % \includegraphics[width=0.32\textwidth]{figures/diffusion_step_num_ablation/val_loss-diffuse_steps.pdf}}
        % %     \subfigure{
        % % \includegraphics[width=0.32\textwidth]{figures/diffusion_step_num_ablation/diffuse_steps-qed.pdf}}
        %     \subfigure{
        % \includegraphics[width=0.32\textwidth]{figures/diffusion_step_num_ablation/diffuse_steps-sa.pdf}}
        %     \subfigure{
        % \includegraphics[width=0.32\textwidth]{figures/diffusion_step_num_ablation/diffuse_steps-vina_score_only.pdf}}
        % %     \subfigure{
        % % \includegraphics[width=0.32\textwidth]{figures/diffusion_step_num_ablation/diffuse_steps-vina_minimize.pdf}}
        % %     \subfigure{
        % % \includegraphics[width=0.32\textwidth]{figures/diffusion_step_num_ablation/diffuse_steps-vina_dock.pdf}}
\caption{Ablation study on diffusion step number. We compare our models with TargetDiff \citep{guan20233d} in terms of best validation loss, QED, SA, Vina Score under different diffusion step number settings.}
\label{fig:diffusion_step_num_ablation}
\end{center}
\end{figure*}

\subsection{Ablation Studies}
% \begin{itemize}
%     \item Training / Sampling Efficiency of DecompDiff
%     \item Ref-prior vs. Beta-prior
%     \item w/ Drift vs. w/o Drift
% \end{itemize}
Since our model is composed of multiple novel designs, including decomposed priors, bond diffusion and additional guidance, we perform comprehensive ablation studies to verify our hypothesis on the effects of each design.

\begin{figure}[t]
\begin{center}
\centerline{
    \includegraphics[width=0.45\textwidth]      {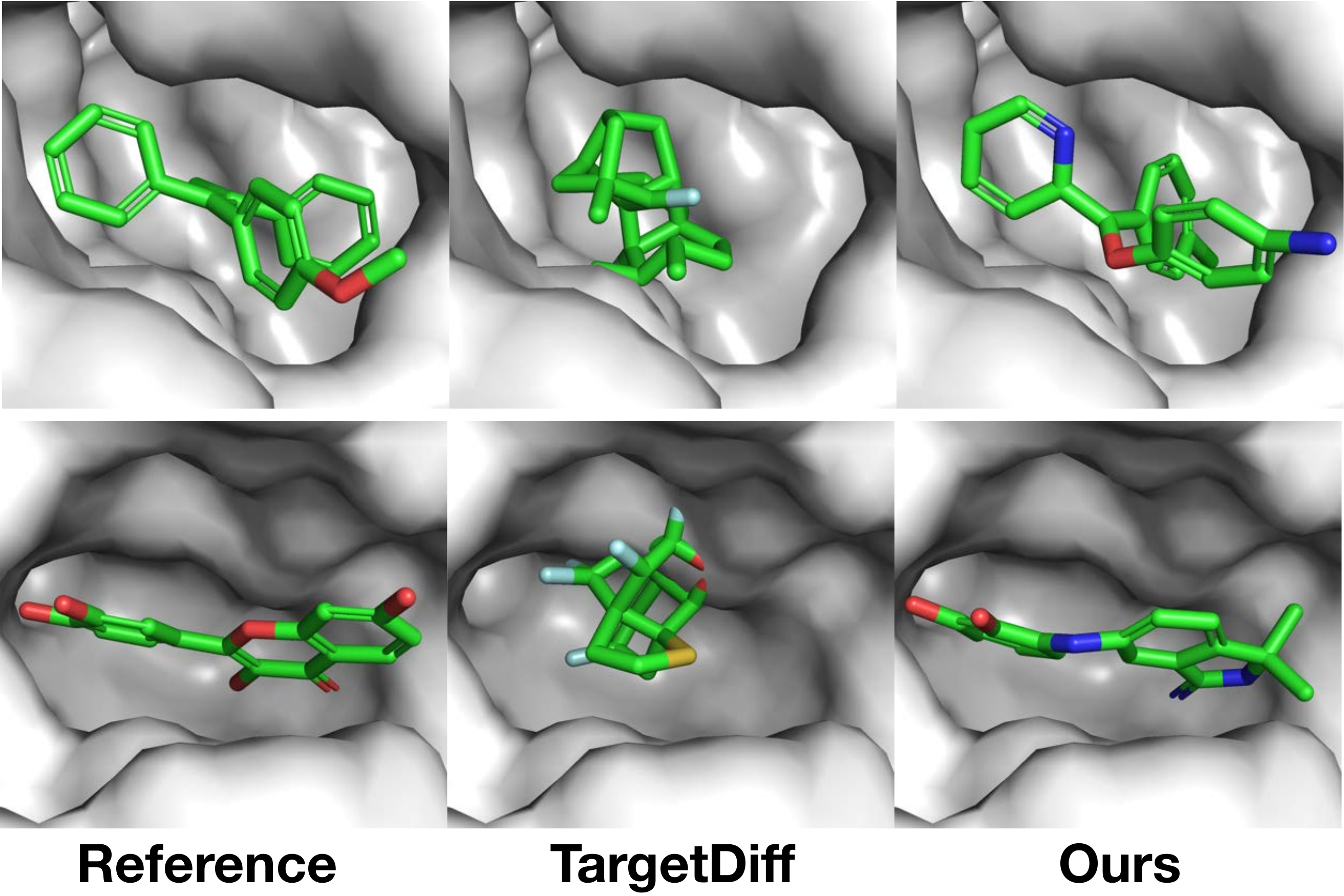}
}
\caption{Visualization of reference binding molecules (left column), molecules generated by TargetDiff \citep{guan20233d} (middle column), and our model (right column) with only $200$ sampling steps on protein 4H3C (top row) and 2F2C (bottom row).}
\label{fig:diffusion_step_num_example}
\end{center}
\vspace{-4mm}
\end{figure}

\paragraph{Effect of Decomposition and Prior} Our primary hypothesis is that decomposing the drug space with prior knowledge can improve the training and sampling efficiency, and thus boosting the molecular generation performance. 
To verify it, we compare our model with TargetDiff \citep{guan20233d} under different number of diffusion steps. We first train our model and TargetDiff with $\{200, 400, 600, 800, 1000\}$ diffusion steps and evaluate the generated molecules by sampling the same number of steps as training on 16 selected pockets with clear decomposition. As shown in \cref{fig:diffusion_step_num_ablation}, our model can achieve better validation loss under each setting. Under the same setting, the validation loss can be viewed as a surrogate of negative Evidence Lower Bound (ELBO) and lower validation loss means the model can better approximate the data distribution. The fact that the model trained with fewer diffusion steps achieves lower validation loss is because it fits noises better at fewer time steps with limited model capacity. \cref{fig:diffusion_step_num_ablation} also shows our model can generate high-quality ligand molecules (high QED and SA, low Vina) even with fewer sampling steps. \cref{fig:diffusion_step_num_example} shows examples of ligand molecules generated by sampling only $200$ steps. With limited sampling steps, our model can already generate rational molecules, while TargetDiff tends to generate unrealistic local structures, such as messy rings.

Besides the training/sampling efficiency analysis, we also investigate the influence of different priors. We mainly explore two kinds of priors: \textit{Ref Prior} is estimated from the reference molecule with a Gaussian distribution through maximum likelihood estimation. \textit{Pocket Prior} utilizes the subpockets within the target binding site extracted by AlphaSpace2 \citep{rooklin2015alphaspace} to estimate the prior center and a neural classifier to estimate the number of ligand atoms and prior standard deviation. See more details in Appendix \ref{sec:prior_generation}. In Table \ref{tab:prior_abl}, we show the affinity-related metrics of an atom-only version (no bond diffusion) of our model with different priors. 
Compared with TargetDiff, a baseline without decomposition and informative prior, our model can generate better binding molecules with appropriate prior knowledge.
Molecules generated using \textit{Pocket Prior} achieve better results compared to \textit{Ref Prior}, which could be caused by the fact that the reference molecule may not be ideal for the target and directly extract knowledge prior from the target binding site could be a better choice.
% in these cases. 

\begin{table}[t]
\vspace{-2mm}
    \tabcolsep 3pt
    \centering
    \caption{The influence of decomposed prior.}
    % \vspace{2mm}
    \begin{adjustbox}{width=0.5\textwidth}
    % \renewcommand{\arraystretch}{1.4}
% \begin{tabular}{l|cc|cc|cc|cc|cc|cc|c}
% \toprule
% \diagbox{Model}{Metric} & \multicolumn{2}{c|}{Vina Score ($\downarrow$)} & \multicolumn{2}{c|}{Vina Min ($\downarrow$)} & \multicolumn{2}{c|}{Vina Dock ($\downarrow$)} & \multicolumn{2}{c|}{High Affinity ($\uparrow$)} & \multicolumn{2}{c|}{QED ($\uparrow$)}   & \multicolumn{2}{c|}{SA ($\uparrow$)} & Success Rate ($\uparrow$)  \\
%  & Avg. & Med. & Avg. & Med. & Avg. & Med. & Avg. & Med. & Avg. & Med. & Avg. & Med. & Avg.  \\
% \midrule

% Ours* - Standard Prior & -5.32 & -5.99 & -6.42 & -6.51 & -6.42 & -6.51 & 50.9\% & 60.4\% & 0.49 & 0.50 & 0.61 & 0.60 &  \\

% % DecompDiff - Base &  -4.747 & -5.921 & -6.171 & -6.474 & * & * & * & * & 0.506 & 0.521 & 0.633 & 0.610 & * & * \\

% Ours - Ref Prior & \textbf{-5.45} & -5.57 & -6.19 & -6.16 & -7.16 & -7.19 & 50.9\% & 49.7\% & 0.50 & 0.51 & 0.66 & 0.64 & 15.90\% \\

% % DecompDiff - Beta & -2.73 & -3.20 & -6.43 & -6.96 & * & * & * & * & 0.506 & 0.521 & 0.633 & 0.610 & * & * \\

% Ours - Beta Prior & -5.17 & \textbf{-6.62} & \textbf{-7.13} & \textbf{-7.60} & -8.62 & -8.82 & \textbf{73.0\%} & \textbf{92.5\%} & 0.37 & 0.34 & 0.55 & 0.56 & 15.92\% \\

% % Reference   & -6.36 & -6.46 & -6.71 & -6.49 & -7.45 & -7.26 & -  & - & 0.48 & 0.47 & 0.73 & 0.74 & * & *    \\

% \bottomrule
% \end{tabular}
% \renewcommand{\arraystretch}{1}

\renewcommand{\arraystretch}{1.}
\begin{tabular}{l|cc|cc|cc}
\toprule
% \diagbox{Model}{Metric} 
\multirow{2}{*}{Methods} & \multicolumn{2}{c|}{Vina Score ($\downarrow$)} & \multicolumn{2}{c|}{Vina Min ($\downarrow$)}  & \multicolumn{2}{c}{Vina Dock ($\downarrow$)} \\
 & Avg. & Med. & Avg. & Med. & Avg. & Med.  \\
\midrule

% TargetDiff-Ref      & -5.32 & -5.99 & -6.42 & -6.51 &  &  \\
TargetDiff          & -5.47 & -6.30 & -6.64 & -6.83 & -7.80 & -7.91 \\
Ours - Ref Prior    & \textbf{-6.11} & -6.16 & -6.68 & -6.58 & -7.53 & -7.59 \\
Ours - Pocket Prior & -5.72 & \textbf{-7.49} & \textbf{-7.66} & \textbf{-8.33} & \textbf{-9.08} & \textbf{-9.33} \\

\bottomrule
\end{tabular}
\renewcommand{\arraystretch}{1}

    \end{adjustbox}
    \label{tab:prior_abl}
    \vspace{-5mm}
\end{table}

\begin{table}[t]
    \vspace{-2mm}
    \centering
    \caption{The influence of bond diffusion.}
    % \vspace{2mm}
    \begin{adjustbox}{width=0.5\textwidth}
    % \renewcommand{\arraystretch}{1.2}
% \begin{tabular}{l|cc|cc|cc|cc|cc|cc|c}
% \toprule
% \diagbox{Model}{Metric} & \multicolumn{2}{c|}{Vina Score ($\downarrow$)} & \multicolumn{2}{c|}{Vina Min ($\downarrow$)} & \multicolumn{2}{c|}{Vina Dock ($\downarrow$)} & \multicolumn{2}{c|}{High Affinity ($\uparrow$)} & \multicolumn{2}{c|}{QED ($\uparrow$)}   & \multicolumn{2}{c|}{SA ($\uparrow$)} & Success Rate ($\uparrow$)  \\
%  & Avg. & Med. & Avg. & Med. & Avg. & Med. & Avg. & Med. & Avg. & Med. & Avg. & Med. & Avg.  \\
% \midrule

% DecompDiff - Atom & -6.11 & -6.16 & -6.68 & -6.58 & -7.53 & -7.59 & 50.9\% & 60.4\% & 0.48 & 0.48 & 0.61 & 0.60 & 11.21\% \\

% % DecompDiff - Base &  -4.747 & -5.921 & -6.171 & -6.474 & * & * & * & * & 0.506 & 0.521 & 0.633 & 0.610 & * & * \\

% DecompDiff - Bond & -5.45 & -5.57 & -6.19 & -6.16 & -7.16 & -7.19 & 50.9\% & 49.7\% & 0.50 & 0.51 & 0.66 & 0.64 & 15.90\% \\

% % DecompDiff - Beta & -2.73 & -3.20 & -6.43 & -6.96 & * & * & * & * & 0.506 & 0.521 & 0.633 & 0.610 & * & * \\

% % DecompDiff - Beta Prior & -5.17 & \textbf{-6.62} & \textbf{-7.13} & \textbf{-7.60} & -8.62 & -8.82 & \textbf{73.0\%} & \textbf{92.5\%} & 0.37 & 0.34 & 0.55 & 0.56 & 15.92\% \\

% % Reference   & -6.36 & -6.46 & -6.71 & -6.49 & -7.45 & -7.26 & -  & - & 0.48 & 0.47 & 0.73 & 0.74 & * & *    \\

% \bottomrule
% \end{tabular}
% \renewcommand{\arraystretch}{1}

\renewcommand{\arraystretch}{1.2}
\begin{tabular}{l|cc|cc|cc|cc|cc|cc|c}
\toprule
% \diagbox{Model}
\multirow{2}{*}{Methods} & \multicolumn{2}{c|}{Vina Dock ($\downarrow$)} & \multicolumn{2}{c|}{QED ($\uparrow$)} & \multicolumn{2}{c|}{SA ($\uparrow$)} & Success ($\uparrow$)  \\
 & Avg. & Med. & Avg. & Med. & Avg. & Med. & Avg.  \\
\midrule

Ours - Atom & \textbf{-7.53} & \textbf{-7.59} & 0.48 & 0.48 & 0.61 & 0.60 & 11.21\% \\

% DecompDiff - Base &  -4.747 & -5.921 & -6.171 & -6.474 & * & * & * & * & 0.506 & 0.521 & 0.633 & 0.610 & * & * \\

Ours - Bond & -7.10 & -7.14 & \textbf{0.51} & \textbf{0.51} & \textbf{0.66} & \textbf{0.65} & \textbf{15.38\%} \\

% DecompDiff - Beta & -2.73 & -3.20 & -6.43 & -6.96 & * & * & * & * & 0.506 & 0.521 & 0.633 & 0.610 & * & * \\

% DecompDiff - Beta Prior & -5.17 & \textbf{-6.62} & \textbf{-7.13} & \textbf{-7.60} & -8.62 & -8.82 & \textbf{73.0\%} & \textbf{92.5\%} & 0.37 & 0.34 & 0.55 & 0.56 & 15.92\% \\

% Reference   & -6.36 & -6.46 & -6.71 & -6.49 & -7.45 & -7.26 & -  & - & 0.48 & 0.47 & 0.73 & 0.74 & * & *    \\

\bottomrule
\end{tabular}
\renewcommand{\arraystretch}{1}
    \end{adjustbox}
    \label{tab:bond_diff_abl}
\vspace{-2mm}
\end{table}

\paragraph{Effect of Bond Diffusion}
Our motivation to include bond generation in the model is to reduce the ratio of unrealistic 2D structures caused by inaccurate bond prediction of the post-processing algorithm. 
We see in Table \ref{tab:bond_diff_abl} that bond diffusion substantially improves QED, SA, and Success Rate, indicating simultaneously modeling atoms and bonds can generate more reasonable 2D structures. 

\begin{table}[t]
 % \vspace{-1mm}
    \centering
    \caption{The influence of validity guidance in sampling.}
    % \vspace{2mm}
    \begin{adjustbox}{width=0.5\textwidth}
    \renewcommand{\arraystretch}{1.}
\begin{tabular}{l|cc|cc|cc}
\toprule
% \diagbox{Model}{Metric} 
\multirow{2}{*}{Methods} & \multicolumn{2}{c|}{Complete ($\uparrow$)} & \multicolumn{2}{c|}{Vina Score ($\downarrow$)} & \multicolumn{2}{c}{Vina Min ($\downarrow$)} \\
% & \multicolumn{2}{c|}{QED ($\uparrow$)}   & \multicolumn{2}{c}{SA ($\uparrow$)} \\
 & Avg. & Med. & Avg. & Med. & Avg. & Med. \\
 % & Avg. & Med. & Avg. & Med. \\
\midrule

TargetDiff - Ref & 0.91 & 0.96 & -5.32 & -5.99 & -6.42 & -6.51 \\
% & 0.49 & 0.50 & 0.61 & 0.60 \\

Ours - No Drift & 0.89 & 0.95 & -4.75 & -5.92 & -6.17 & -6.47 \\
% & 0.51 & 0.52 & 0.63 & 0.61\\

Ours - ArmSca Drift & 0.94 & 0.98 & -4.84 & -5.99 & -6.20 & -6.50 \\
% & 0.50 & 0.51 & 0.63 & 0.61 \\

Ours - Clash Drift & 0.87 & 0.94 & -5.79 & -6.00 & -6.58 & -6.54 \\
% & \textbf{0.51} & \textbf{0.52} & \textbf{0.63} & \textbf{0.62} \\

Ours - All Drift & \textbf{0.94} & \textbf{0.98} & \textbf{-6.11} & \textbf{-6.16} & \textbf{-6.68} & \textbf{-6.58} \\
% & 0.48 & 0.49 & 0.61 & 0.61  \\

% Reference   & -6.36 & -6.46 & -6.71 & -6.49 & 0.48 & 0.47 & 0.73 & 0.74 \\

\bottomrule
\end{tabular}

    \end{adjustbox}
    \label{tab:drift_abl}
\vspace{-4mm}
\end{table}

\paragraph{Effect of Validity Guidance}
To show the effectiveness of validity guidance proposed in Sec. \ref{sec:guidance}, we test different guidance drifts during the sampling phase in our decomposed atom diffusion model. As shown in Table \ref{tab:drift_abl}, the arm-scaffold connection drift and the clash drift have a positive influence on the complete rate and Vina score respectively. Combining them (All Drift) achieves even better performance in both complete rate and Vina score.

\section{Conclusions}
In this work, we proposed \method for SBDD by decomposing the drug space. The introduction of informative priors to the diffusion model improves the training and sampling efficiency and significantly improves the binding affinity measured by Vina. We also extend the 
% traditional diffusion models operating on 3D point clouds
diffusion model to simultaneously generate atom coordinates, atom types, and bond types, which is beneficial to generate more realistic molecules. Validity guidance during sampling is simple but effective under our decomposition framework. 
%  Future work

\section*{Acknowledgements}
We thank all the reviewers for their feedbacks through out the review cycles of the manuscript. This work is supported by National Natural Science Foundation of China (62206291, 62141608, 62236010), National Key R\&D Program of China No.
2021YFF1201600, U.S. National Science Foundation under grant no. 2019897 and U.S. Department of Energy award DE-SC0018420.

\newpage
\bibliography{main}
\bibliographystyle{icml2023}

%%%%%%%%%%%%%%%%%%%%%%%%%%%%%%%%%%%%%%%%%%%%%%%%%%%%%%%%%%%%%%%%%%%%%%%%%%%%%%%
%%%%%%%%%%%%%%%%%%%%%%%%%%%%%%%%%%%%%%%%%%%%%%%%%%%%%%%%%%%%%%%%%%%%%%%%%%%%%%%
% APPENDIX
%%%%%%%%%%%%%%%%%%%%%%%%%%%%%%%%%%%%%%%%%%%%%%%%%%%%%%%%%%%%%%%%%%%%%%%%%%%%%%%
%%%%%%%%%%%%%%%%%%%%%%%%%%%%%%%%%%%%%%%%%%%%%%%%%%%%%%%%%%%%%%%%%%%%%%%%%%%%%%%

\appendix
\onecolumn

\section{The Generation Details of Pocket Prior Centers}
\label{sec:prior_generation}

\begin{figure}[h]
\begin{center}
    \subfigure[Alpha-atoms.]{
        \label{alpha-atoms} %% label
        \includegraphics[width=0.31\textwidth]{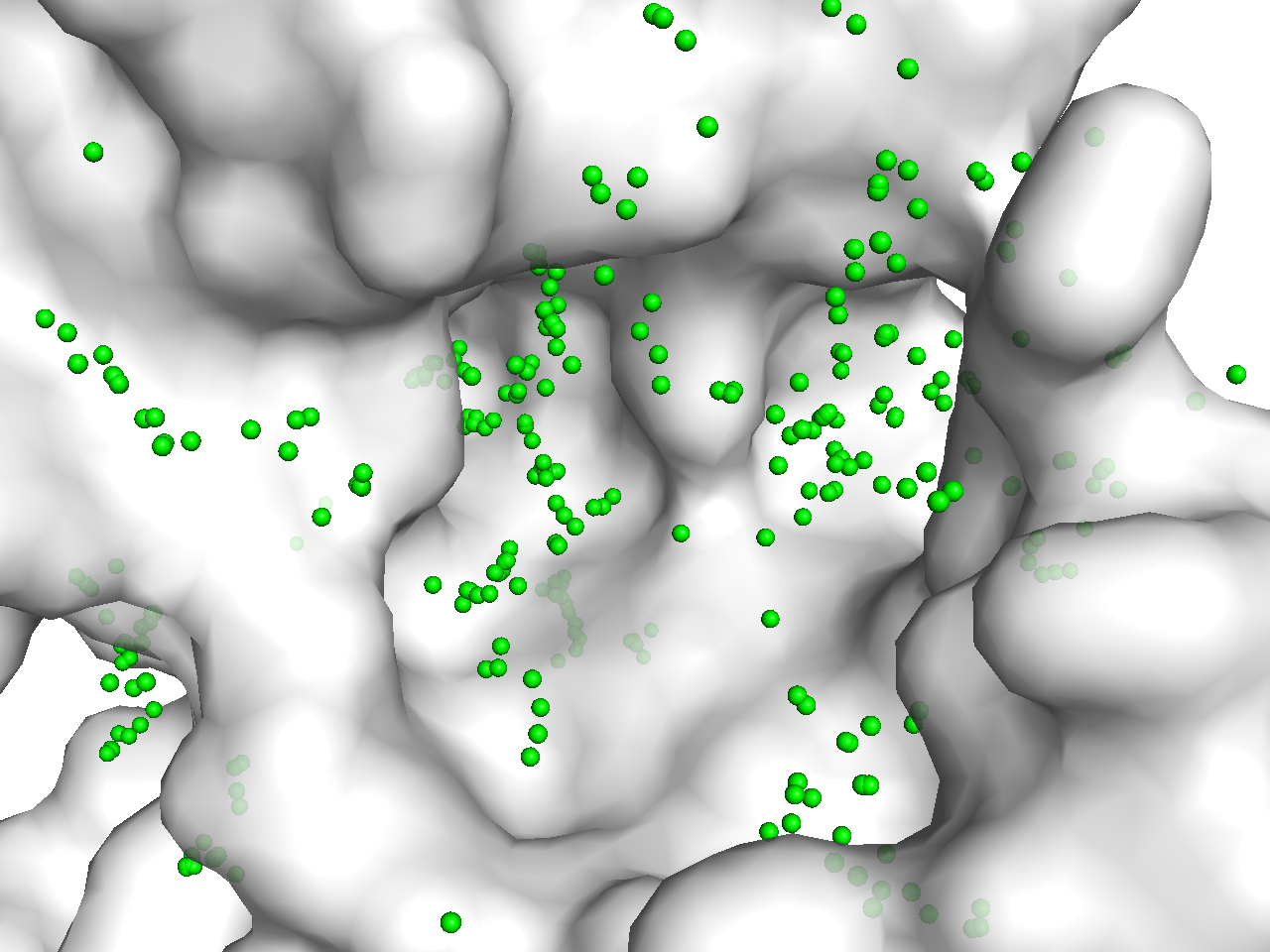}}
    \subfigure[Beta-atoms.]{
        \label{beta-atoms} %% label 
        \includegraphics[width=0.31\textwidth]{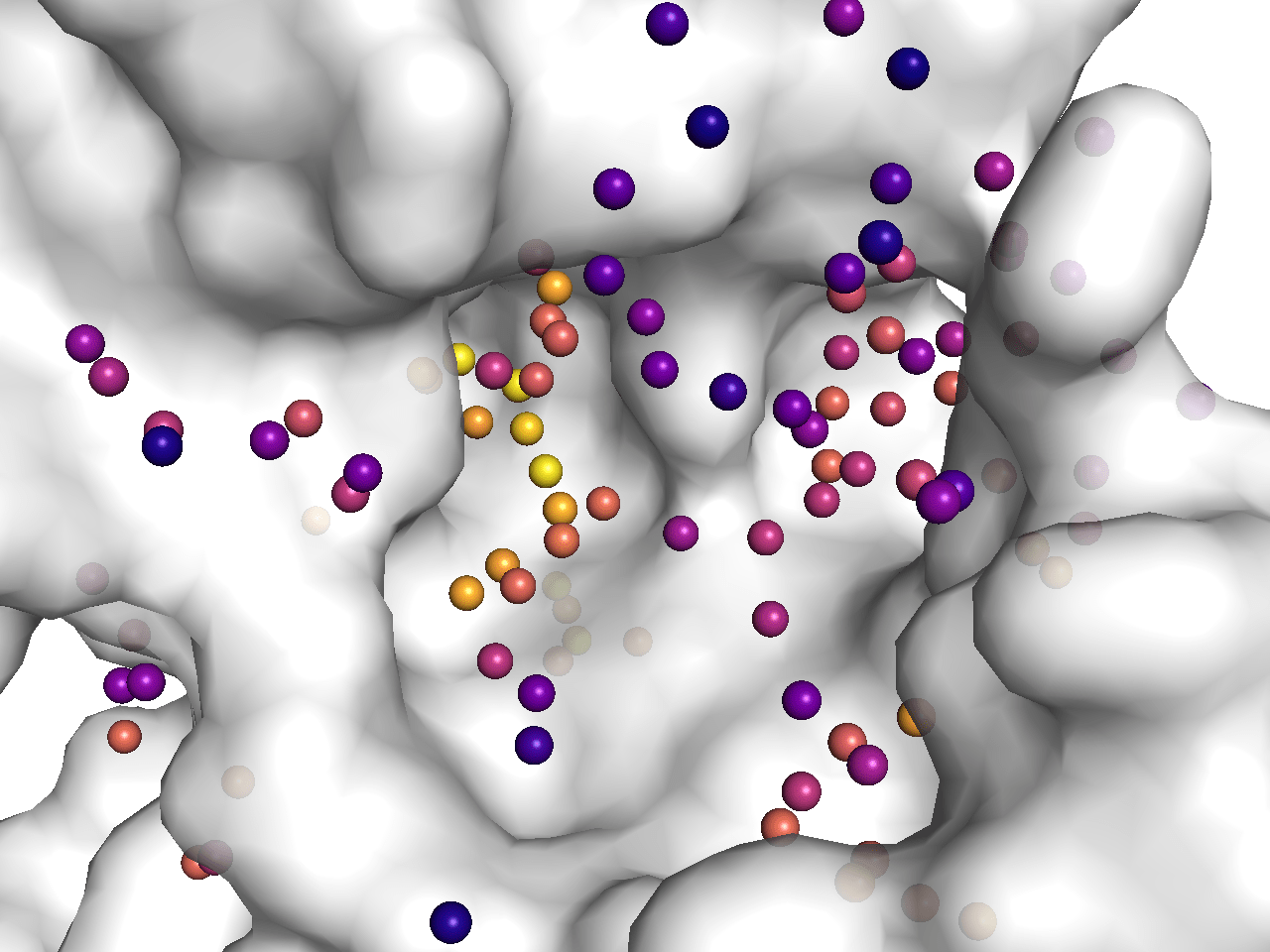}}
    \subfigure[Alpha-graph.]{
        \label{alpha-graph} %% label 
        \includegraphics[width=0.31\textwidth]{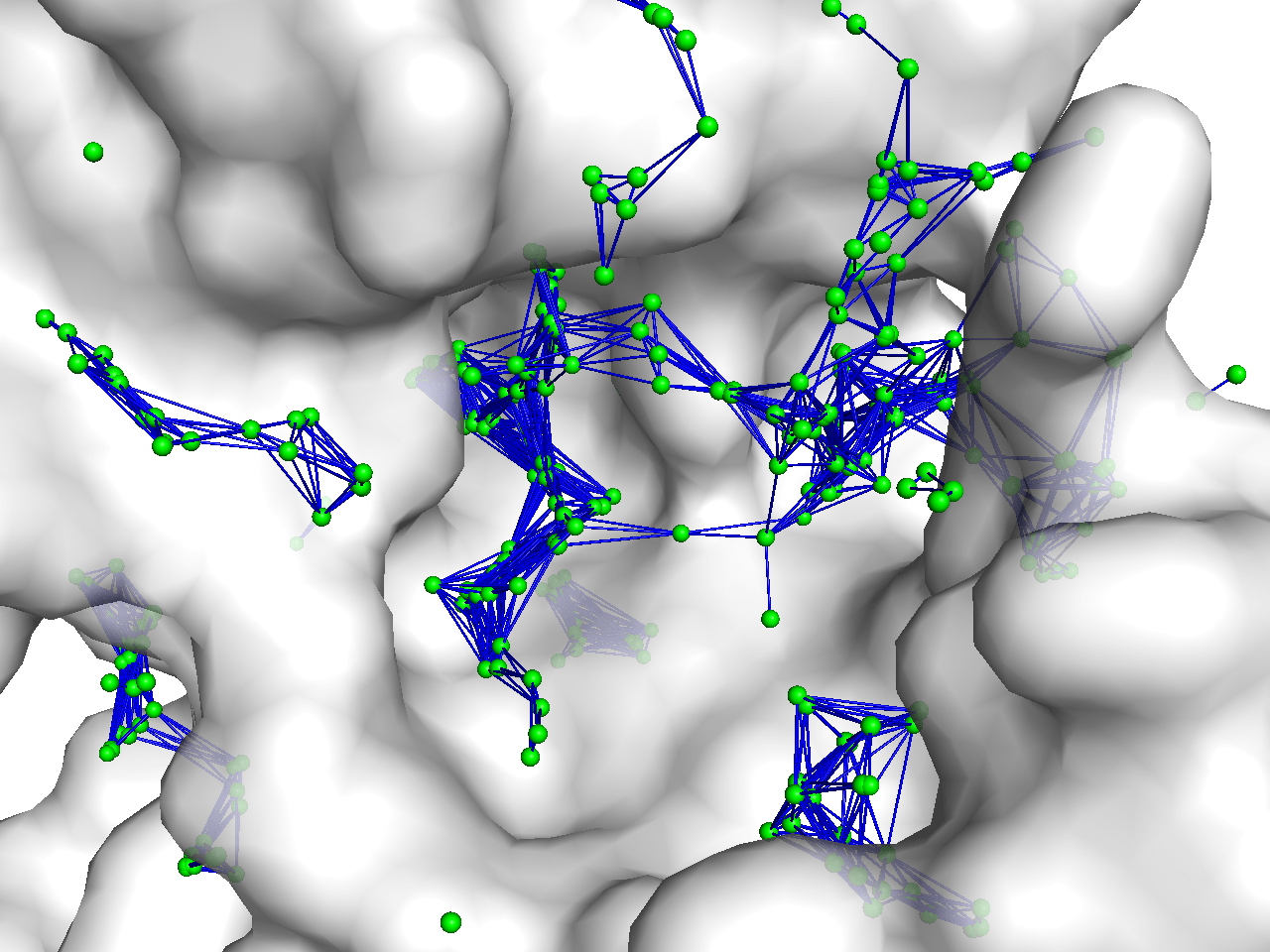}}
    \subfigure[Beta-clusters.]{
        \label{beta-cluster} %% label 
        \includegraphics[width=0.31\textwidth]{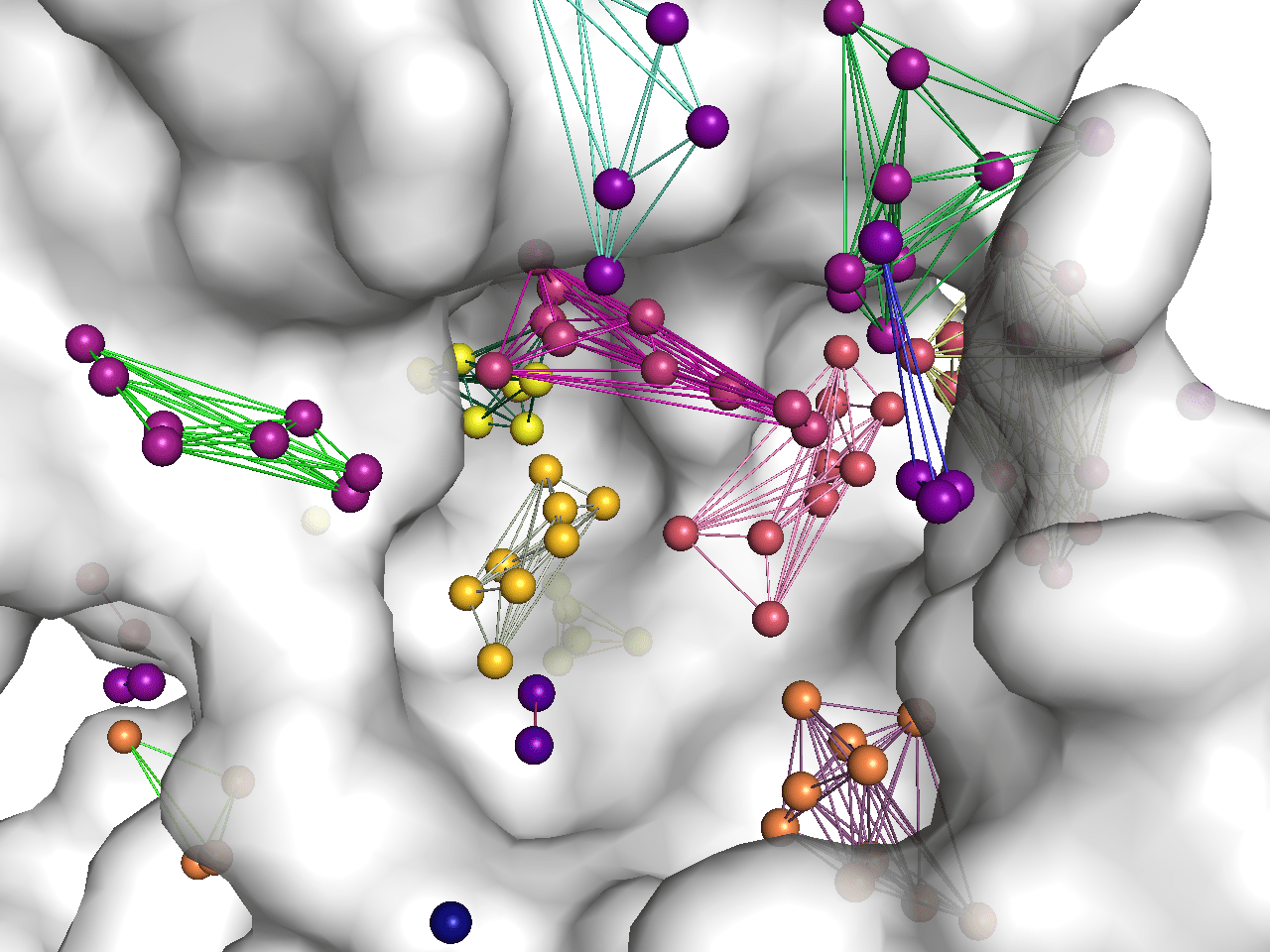}}
    \subfigure[Filtered beta-clusters.]{
        \label{beta-cluster-final} %% label 
        \includegraphics[width=0.31\textwidth]{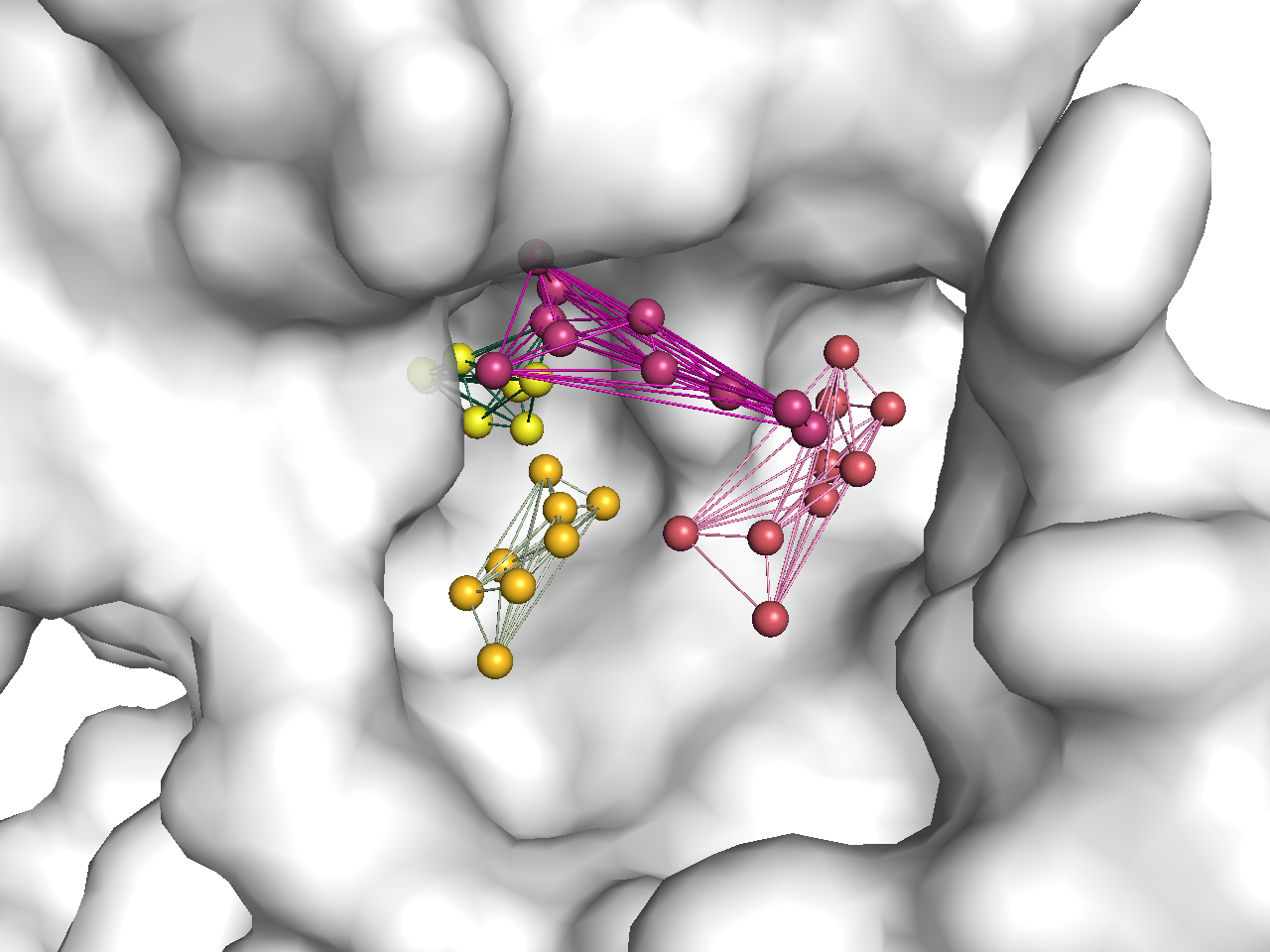}}
    \subfigure[Priors.]{
        \label{priors-appendix} %% label 
        \includegraphics[width=0.31\textwidth]{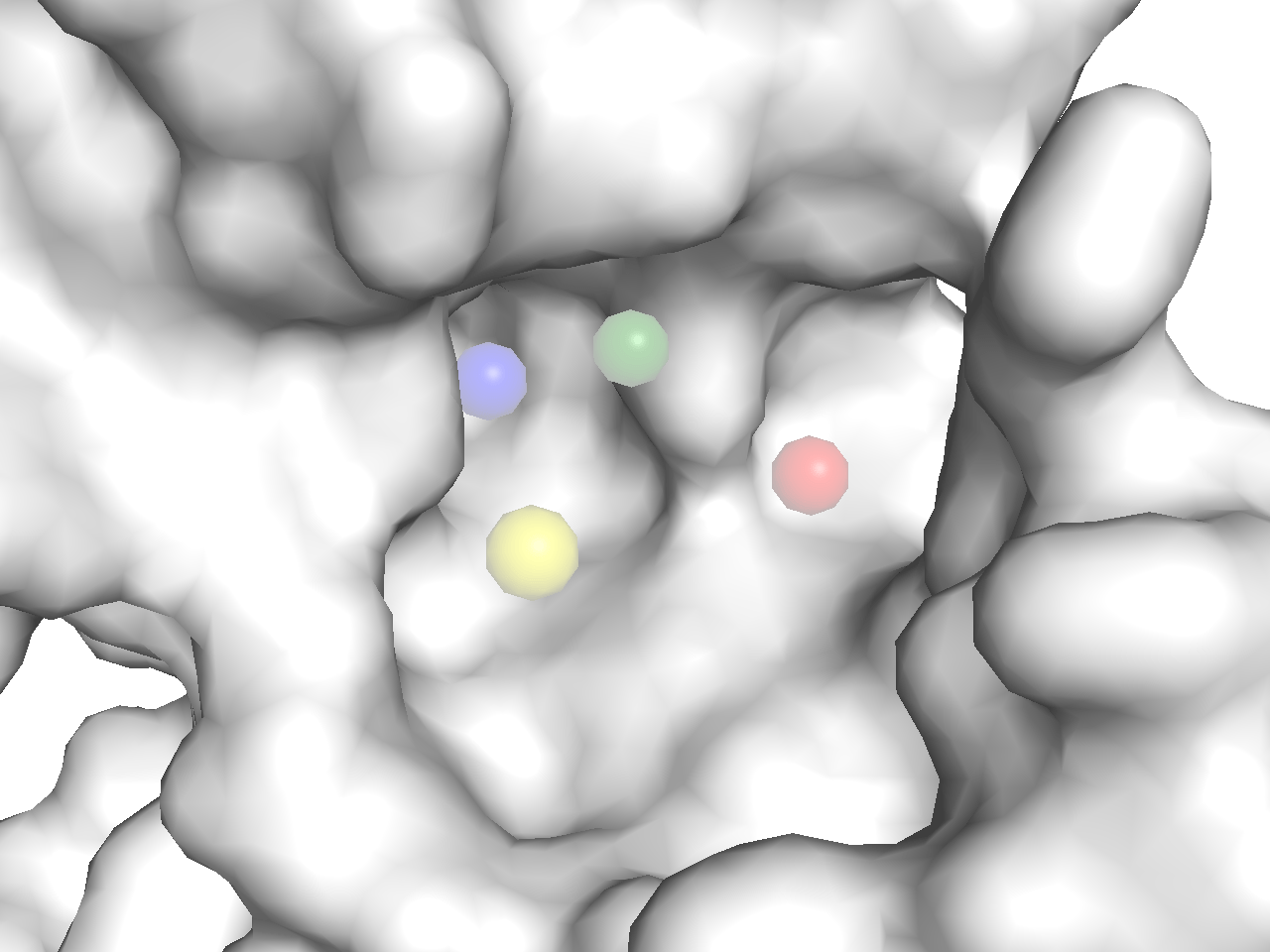}}
    \caption{Illustration of prior centers generation process. (a) Alpha-atoms. (b) Beta-atoms with beta-scores. The closer the hue of a beta-atom to yellow, the better its beta-score is. (c) Alpha-graph. Here we only visualize the alpha-atoms and omit beta-cluster centers for clarity. (d) Beta-clusters. Each beta-cluster forms a fully-connected graph consisting of its beta-atoms. All beta-atoms in a beta-cluster are visualized in the same color, corresponding to the score. (e) Filtered beta-clusters, and corresponding (f) Priors. 
    The scaffold prior is shown in yellow, and the arm priors are shown in other colors.} 
\end{center}
\end{figure}

% In this section, we present the procedure for generating prior centers and the related hyperparameters in detail.
Given protein pockets, the arms are the main components of a ligand interacting with pockets, and the scaffold connects the arms to form a ligand.
We obtain the prior centers of arms and scaffolds of ligands in the following steps:
\begin{compactenum}
\item \textbf{Searching alpha-atoms and beta-atoms}: Following \citet{rooklin2015alphaspace,katigbak2020alphaspace}, we search the alpha-atoms and beta-atoms by \texttt{AlphaSpace2}\footnote{\url{https://yzhang.hpc.nyu.edu/AlphaSpace2/}} toolkit.
% which can detect potential pockets on protein surfaces.
An alpha-atom can be viewed as a virtual point, geometrically contacted with the local region of the protein surface~(see \cref{alpha-atoms}). 
And a beta-atom is an alpha-atom group and assigned a beta-score to indicate their pocket ligandability~(see \cref{beta-atoms}). 
Notice that we only consider alpha-atoms and beta-atoms within $10$\AA\ of the reference ligand or the interested region. 
% All alpha-atoms and beta-atoms inside the protein surface are removed. 

\item \textbf{Clustering beta-atoms to beta-clusters}: 
We apply hierarchical clustering on beta-atoms to further group them into beta-clusters, based on the pairwise distances, as shown in \cref{beta-cluster}. 
A pair of beta-atoms whose distance is less than a certain distance (in practice, we use $5.5$\AA) are assigned to the same beta-cluster. 
The score of the beta-cluster is defined as the average beta-scores of the beta-atoms that belong to it. 

\item \textbf{Assigning beta-clusters into arms and scaffolds}:
We filter beta-clusters and determine the scaffold and arm priors with \cref{alg:gen_prior}. 
Specifically, we first find the region contains a proper number of beta-clusters with promising beta-scores.
We then select centers of beta-clusters, according to their contribution to binding affinity and the positions holding scaffold-arm geometric connection, as centers of arm and scaffold priors, respectively. 
% Examples of generated priors are shown in \cref{priors_example}. 
\end{compactenum}
Note that the distance for filtering beta-clusters is not Euclidean in 3D space. 
Instead, we define the distance by introducing a radius graph (called alpha-graph, see~\cref{alpha-graph}) where each alpha-atom is connected to the other alpha-atoms and beta-cluster centers. 
Then, we use the graph geodesic on the alpha-graph between two beta-cluster centers to define their distance (If two beta-cluster centers are unconnected, the distance is defined as $+\infty$). 

% \begin{figure}[ht]
% \vskip 0.2in
% \begin{center}
%     \subfigure{
%         \includegraphics[width=0.31\textwidth]{figures/appendix/prior_generation/prior_data_id_000.png}}
%          \subfigure{
%         \includegraphics[width=0.31\textwidth]{figures/appendix/prior_generation/prior_data_id_001.png}}
%             \subfigure{
%         \includegraphics[width=0.31\textwidth]{figures/appendix/prior_generation/prior_data_id_002.png}}
%             \subfigure{
%         \includegraphics[width=0.31\textwidth]{figures/appendix/prior_generation/prior_data_id_003.png}}
%             \subfigure{
%         \includegraphics[width=0.31\textwidth]{figures/appendix/prior_generation/prior_data_id_005.png}}
%             \subfigure{
%         \includegraphics[width=0.31\textwidth]{figures/appendix/prior_generation/prior_data_id_006.png}}
% \caption{Examples of generated priors.}
% \label{priors_example}
% \end{center}
% \vskip -0.2in
% \end{figure}

 \begin{algorithm}[ht]
   \caption{Procedure of filtering beta-clusters and determining scaffold and arm priors.}
   \label{alg:gen_prior}
\begin{algorithmic}[1]
   \REQUIRE a list of beta-clusters \texttt{clusterList}, scores of the beta-clusters in the list \texttt{score}, maximum number of arms $N$, maximum distance between arms $\delta$, maximum distance between the scaffold and arms $\sigma$, and minimum size of interested connected components $M$.
   \ENSURE list of beta-clusters \texttt{armClusterList} which correspond to the arm priors, and beta-cluster \texttt{scaffoldCluster} which corresponds to the scaffold prior.
   \STATE Compute pairwise distances of centers of beta-clusters in \texttt{clusterList}, and build a graph, in which two clusters are connected if their distance is less than $\delta$.
   \STATE Remove all clusters with connected components constituting less than $M$ nodes from \texttt{clusterList}.
   \STATE Generate a set of clusters \texttt{tempClusterList} with distances less than $\delta$ to the best cluster (the cluster with the highest score).
   \STATE Use the first $N$ clusters in \texttt{tempClusterList} as arm priors \texttt{armClusterList}, and the others as candidates for scaffold priors \texttt{scaffoldClusterList}.
   \STATE Search for the best scaffold cluster \texttt{bestScaffoldCluster} with minimum maximum distance to others in \texttt{scaffoldClusterList}.
   \STATE Ensure the distances between \texttt{bestScaffoldCluster} and clusters in \texttt{armClusterList} are no more than $\sigma$, otherwise assign \texttt{bestScaffoldCluster} = \textsc{NULL}.
   \STATE Assign \texttt{finalClusterList} = \texttt{armClusterList}, and append \texttt{bestScaffoldCluster} to \texttt{finalClusterList}, if $\texttt{bestScaffoldCluster} \ne \textsc{NULL}$.
   \IF{len(\texttt{finalClusterList}) $>2$}
       \STATE \textcolor{teal}{\# When there are at least 3 clusters, search for the centric one as the scaffold prior.}
        \STATE Compute the pairwise distances of centers of clusters in \texttt{finalClusterList}, assign \texttt{scaffoldCluster} as the one with minimum maximum distances to others, and assign \texttt{armClusterList} as the others.
   \ELSE
       \STATE \textcolor{teal}{\# When there are only 2 clusters, the centric one can not be identified.}
        \STATE Assign the cluster with the better \texttt{score} to the only element of \texttt{armClusterList}, and the other one to \texttt{scaffoldCluster}.
   \ENDIF
\end{algorithmic}
\end{algorithm}
\section{The Details of Arms-Scaffold Fragmentation}
\label{sec:fragmentation}
Here, we describe the algorithm for fragmenting a binding molecule into arms and scaffold, given the target protein.
Such fragmentation is used for preparing our training data.

\begin{figure}[ht]
\vskip 0.2in
\begin{center}
    \subfigure{
        \includegraphics[width=0.24\textwidth]{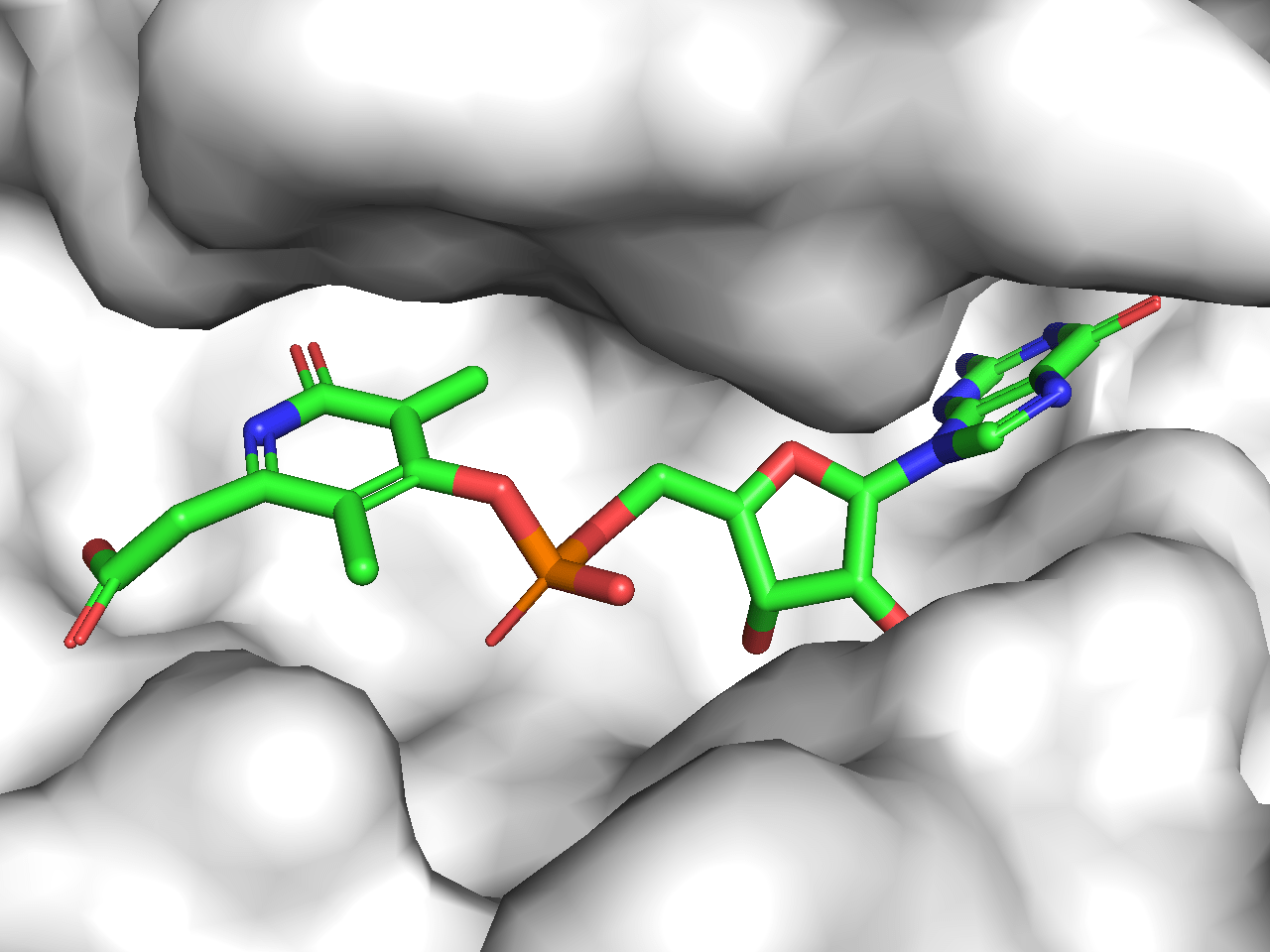}}
    \subfigure{
        \includegraphics[width=0.24\textwidth]{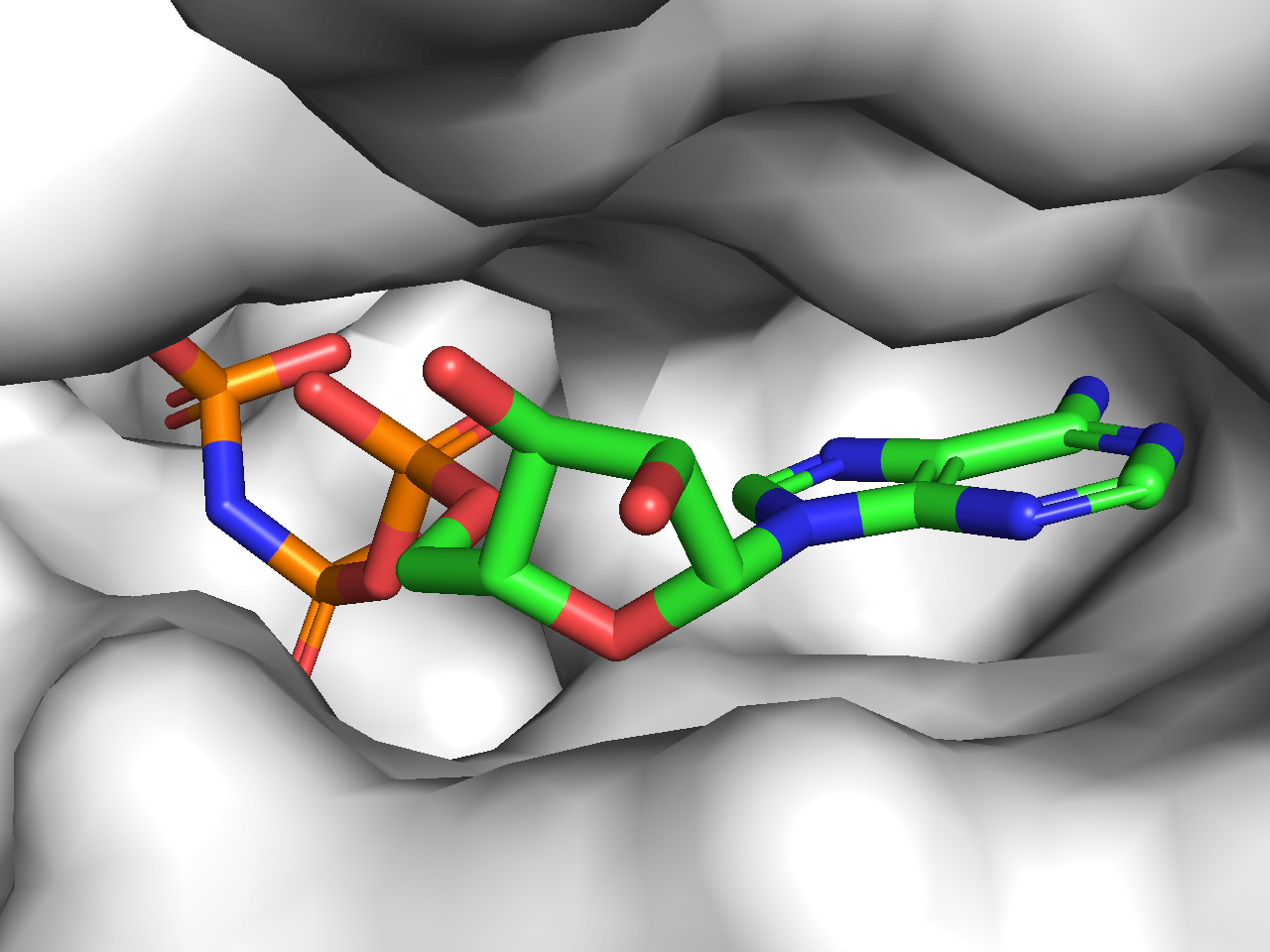}}
    \subfigure{
        \includegraphics[width=0.24\textwidth]{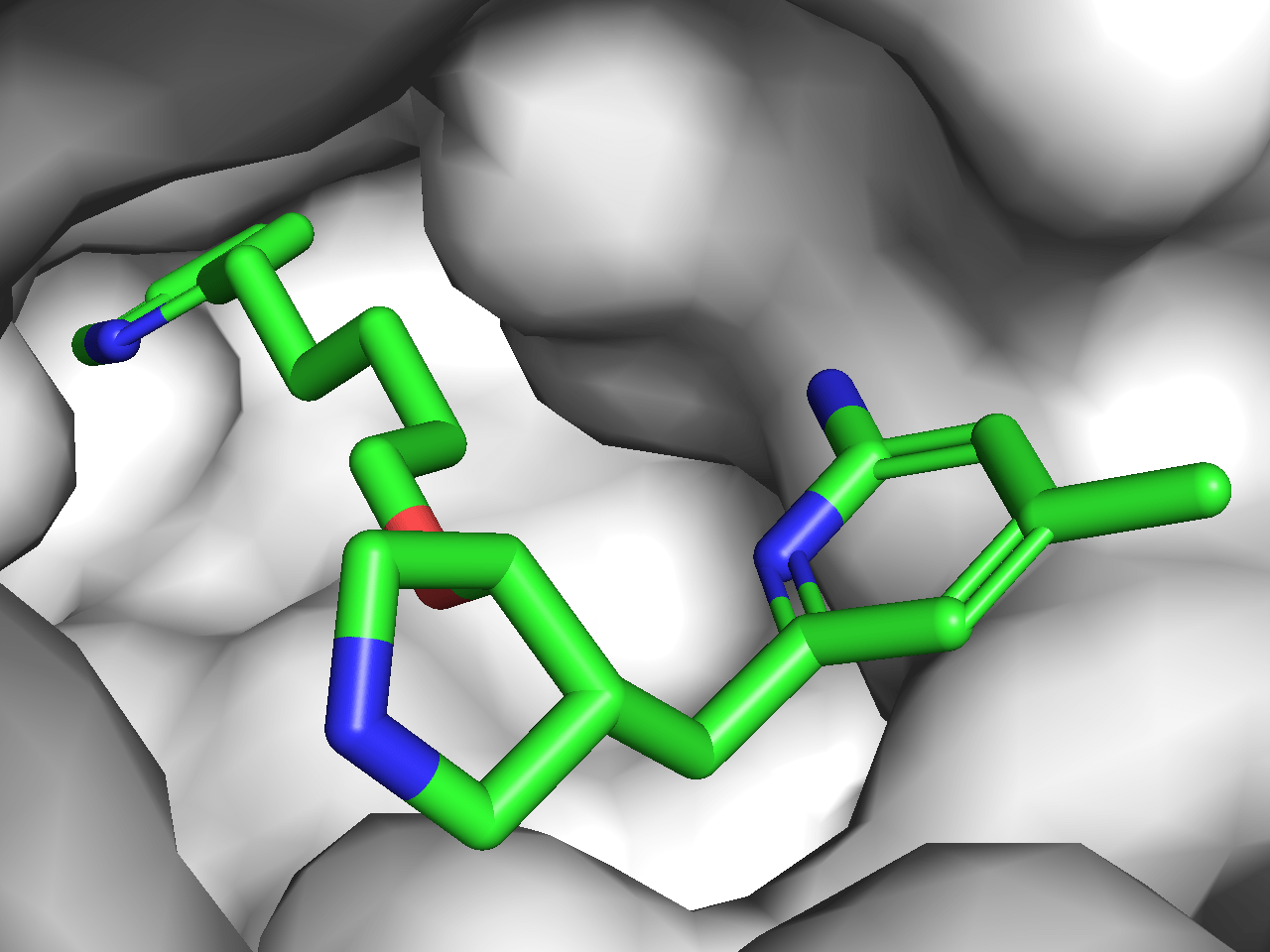}}
    \subfigure{
        \includegraphics[width=0.24\textwidth]{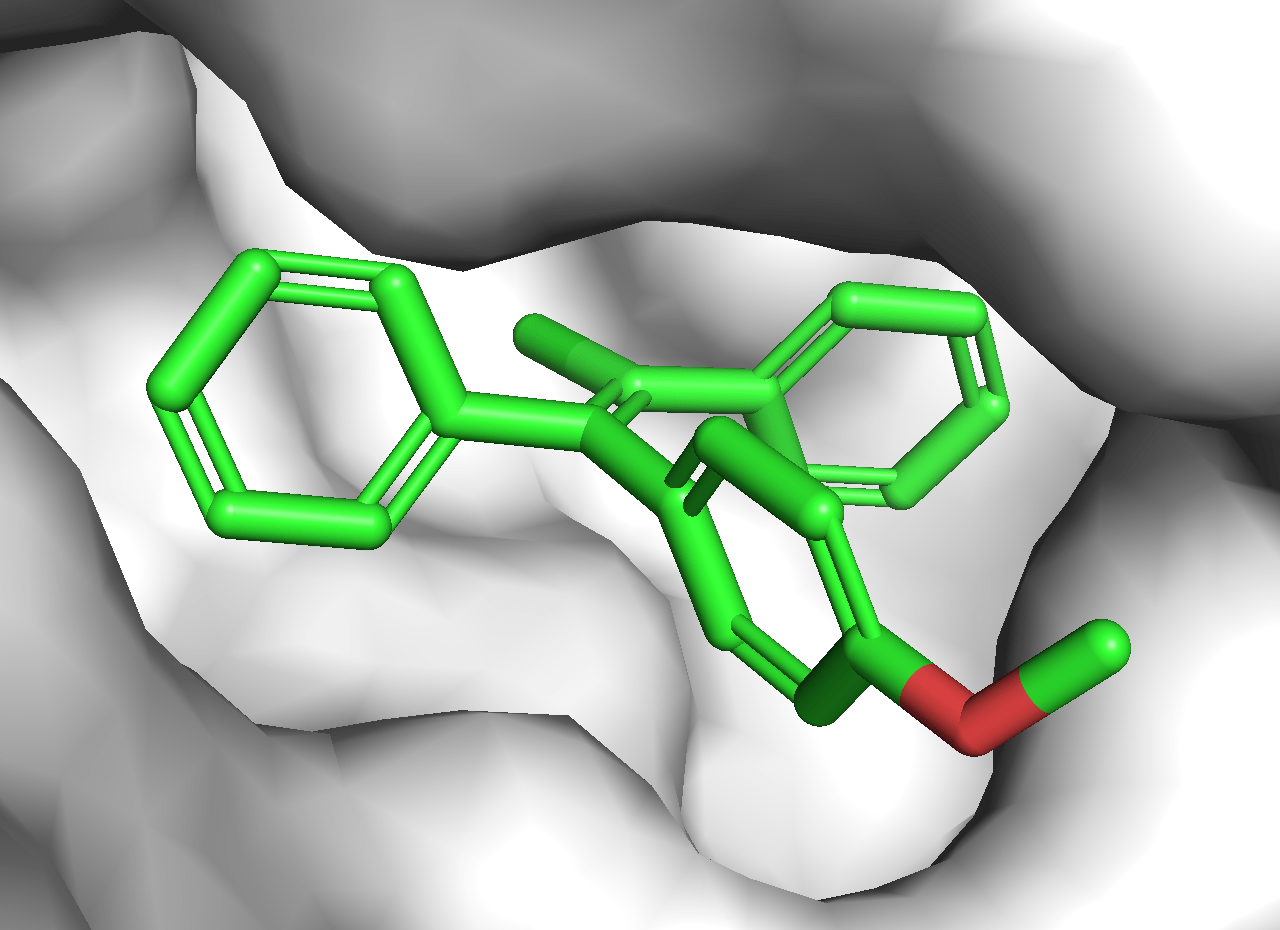}}
    \subfigure{
        \includegraphics[width=0.24\textwidth]{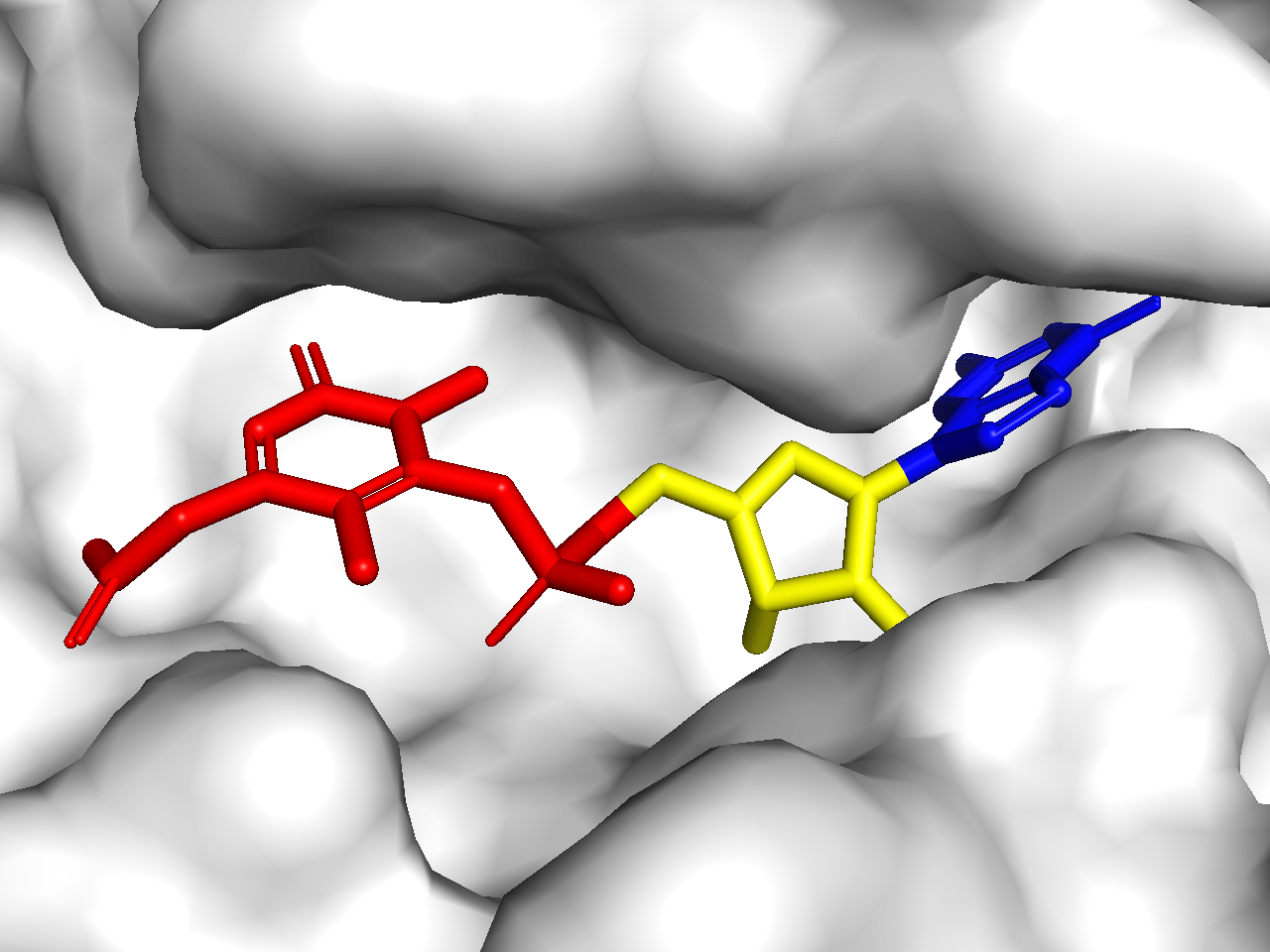}}
    \subfigure{
        \includegraphics[width=0.24\textwidth]{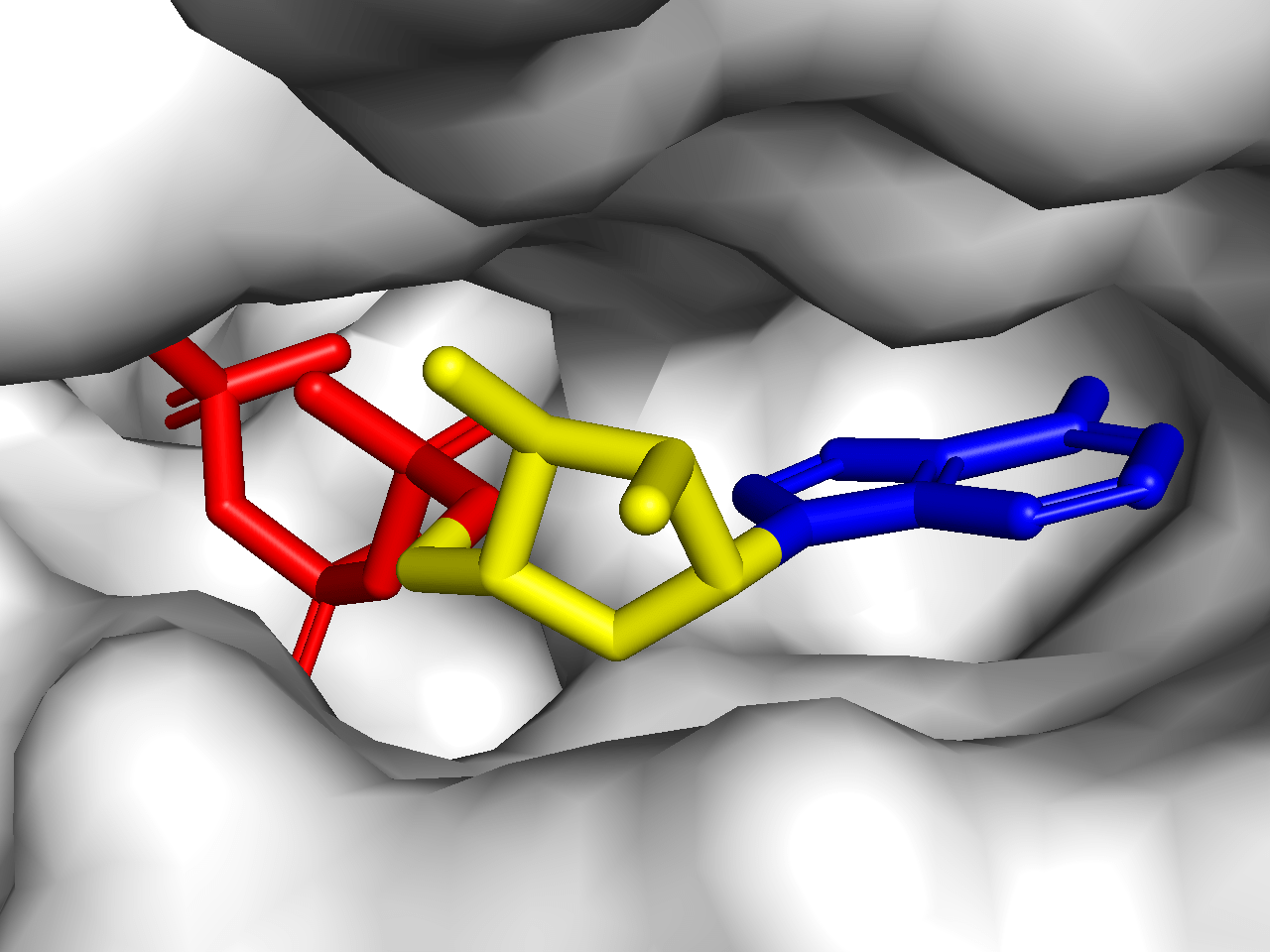}}
    \subfigure{
        \includegraphics[width=0.24\textwidth]{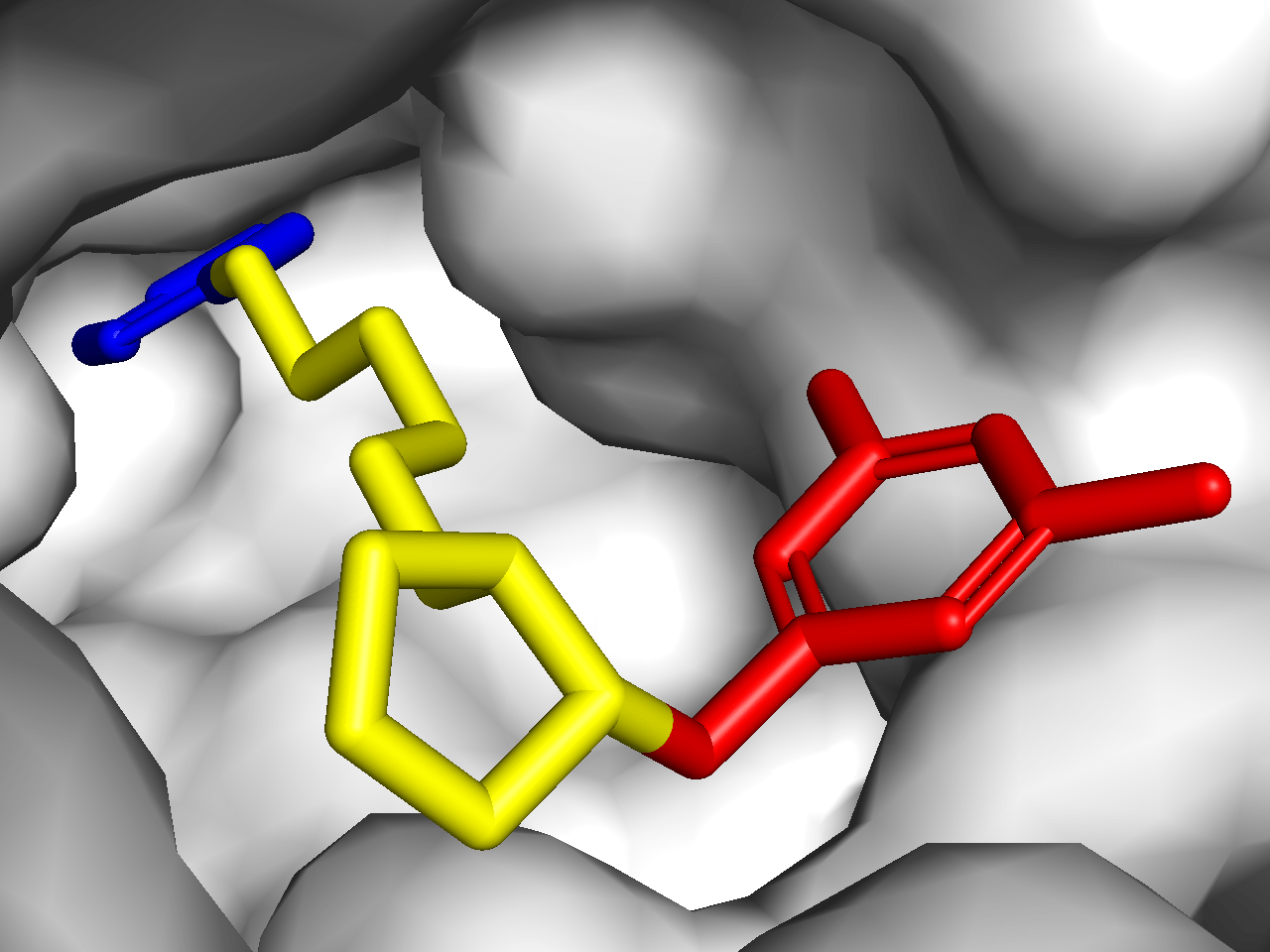}}
    \subfigure{
        \includegraphics[width=0.24\textwidth]{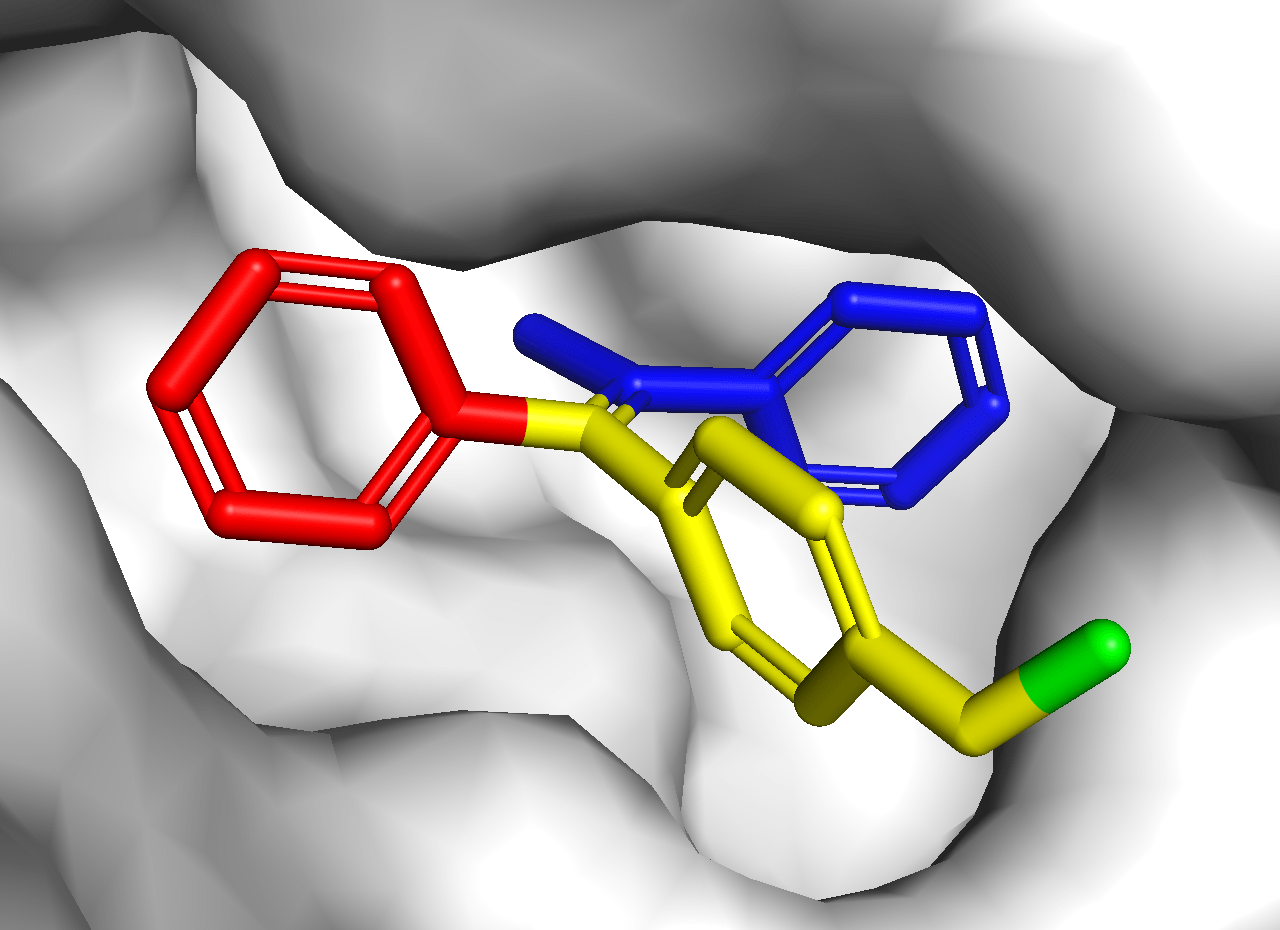}}
\caption{Examples of arms-scaffold fragmentation. Each ligand (top row) is fragmented into arms and a scaffold (bottom row). The scaffold is visualized in yellow. The arms are visualized in other colors.}
\label{fig:fragmentation_example}
\end{center}
% \vskip -0.2in
\end{figure}

We first extract the target protein subpockets with \texttt{AlphaSpace2}~\citep{katigbak2020alphaspace}, which is a surface topographical mapping tool to identify the potential protein binding sites. 
We set the beta atoms' clustering distance as 1.6 \AA, and the pockets' clustering distance as 6.0 \AA. 
These extracted subpockets will be used as the potential arms/scaffold clustering center. 
We then use BRICS \citep{degen2008art} to decompose ligand molecules into fragments. Our goal is to tag these fragments as arms or the scaffold. To achieve this, we perform clustering on these molecular fragments based on the following procedure:
\begin{compactitem}
    \item Perform the linear sum assignment to assign terminal fragments (only one connection site with other molecular parts) to subpockets extracted by AlphaSpace2.
    \item Take centroids of terminal fragments and remaining subpockets (may do not exist) as the \textit{arm} clustering centers. 
    \item Take the centroid of the fragment which is farthest from all existing arm centers as the \textit{scaffold} clustering center. 
    \item Perform nearest neighbor clustering, along which we always make sure the arm fragments are terminal. 
\end{compactitem}
Finally, we can extract arms and scaffold based on the clustering assignment. 
\cref{fig:fragmentation_example} provides examples of our fragmented arms and scaffolds  with their corresponding original ligands and protein pockets.

\newpage
\section{Theoretical Analysis}
\label{sec:proof}
In this section, we will compare our decomposed prior with standard Gaussian prior and analyze its superiority theoretically. The decomposed prior manifests itself in two aspects: (a) It consists of multiple Gaussian distributions with different means rather than a single one; (b) The covariance matrix of each Gaussian distribution does have not to be a unit one. For simplicity, we will show the superiority of the decomposed prior theoretically from these two aspects respectively.

Recall that we denote $\Tilde{\rvx}^{(i)}_{t,k}= \rvx^{(i)}_{t}-\vmu_k$. Additionally, for clarity in the proof, we denote $\Tilde{\rvx}^{(i)}_{t}= \rvx^{(i)}_{t}-\vmu_k = \rvx^{(i)}_{t}-\vmu^{(i)}$ where $k$ satisfies $\eta_{ik}=1$. Similarly, for the standard prior, we denote $\Breve{\rvx}^{(i)}_{t} = \rvx^{(i)}_{t}-\vmu$ here. $\mu_d^{(i)}$ and $\mu_d$ are the $d$th scalar element in $\vmu^{(i)}$ and $\vmu$ respectively.

We first show why the decomposed prior can approximate the data distribution better in aspect (a).
Respecting the natural decomposition of a binding molecule, we make an assumption that $\mathbb{E}[\rvx^{(i)}_{0}] = \vmu_k \in \R^D, \mathbb{E}[\Tilde{\rvx}^{(i)}_{0}\Tilde{\rvx}^{(i)\intercal}_{0}]= \mI \in \R^{D\times D}$ if $\eta_{ik}=1$ and $\mathbb{E}[\Tilde{\rvx}^{(j)}_{0}\Tilde{\rvx}^{(i)\intercal}_{0} ]=\bm{0}, \forall{i\neq j}$. Besides, we assume that the score network is a simple graph neural network with only one layer of linear transformation and summation as an aggregation function, i.e., $\Hat{\rvx}^{(i)}_{0,t}=f_\vtheta(\rvx_t, t)=\sum_{j\neq i}\theta( {\rvx}^{(j)}_{t}-{\rvx}^{(i)}_{t}) + {\rvx}^{(i)}_{t}$ and $\theta = \text{diag}(\theta_1, \theta_2, \cdots, \theta_D) \in \R^{D\times D}$ are learnable parameters constraint to a diagonal matrix with freedom $D$. The diffusion process of the decomposed and standard cases are shown respectively as follows: 
\begin{equation}
\label{eq:appendix_proof_diffuseion_process_decompose}
    \Tilde{\rvx}^{(i)}_{t} = \sqrt{\Bar{\alpha}_t}\Tilde{\rvx}^{(i)}_{0} + \sqrt{1-\Bar{\alpha}_t}\rvepsilon \quad \text{where} \quad
    {\rvx}^{(i)}_{t} = \Tilde{\rvx}^{(i)}_{t} + \vmu^{(i)}
\end{equation} 
\begin{equation}
\label{eq:appendix_proof_diffuseion_process_standard}
    \Breve{\rvx}^{(i)}_{t} = \sqrt{\Bar{\alpha}_t}\Breve{\rvx}^{(i)}_{0} + \sqrt{1-\Bar{\alpha}_t}\rvepsilon \quad \text{where} \quad
    {\rvx}^{(i)}_{t} = \Breve{\rvx}^{(i)}_{t} + \vmu
\end{equation}
where $\rvepsilon\sim \gN(\vzero, \mI)$.

Under the above assumption, the negative evidence lower bound (ELBO) based on the decomposed prior can be derived as: 
\begin{equation}
\begin{aligned}
\label{eq:elbo_decomp_simple}
-\text{ELBO}_{\text{decomp}} 
& = %%%%% 1
\mathbb{E}_{\rvx_0, \rvepsilon}
\left[
\sum_i \sum_k \eta_{ik}
    \left[
    \frac{\Bar{\alpha}_T}{2} 
    \left\Vert \Tilde{\rvx}^{(i)}_{0,k}\right\Vert^2
    +
    \sum_{t=1}^{T-1} \gamma_t 
    \left\Vert {\rvx}^{(i)}_{0,k} - \Hat{\rvx}^{(i)}_{0,t} \right\Vert^2
    \right]
\right] + C
\\ & = %%%%% 2
\mathbb{E}_{\rvx_0, \rvepsilon}
\left[
\sum_k \sum_{i:\eta_{ik}=1}
    \left[
    \frac{\Bar{\alpha}_T}{2} 
    \left\Vert {\rvx}^{(i)}_{0} - \vmu^{(i)}\right\Vert^2
    +
    \sum_{t=1}^{T-1} \gamma_t 
    \left\Vert {\rvx}^{(i)}_{0} - \Hat{\rvx}^{(i)}_{0,t} \right\Vert^2
    \right]
\right] + C
\end{aligned}
\end{equation}
Similarly, the negative ELBO based on the standard Gaussian prior can be derived as: 
\begin{equation}
\begin{aligned}
\label{eq:elbo_standard_simple}
-\text{ELBO}_{\text{standard}} 
& = %%%%% 1
\mathbb{E}_{\rvx_0, \rvepsilon}
\left[
\sum_i \sum_k \eta_{ik}
    \left[
    \frac{\Bar{\alpha}_T}{2} 
    \left\Vert \Breve{\rvx}^{(i)}_{0}\right\Vert^2
    +
    \sum_{t=1}^{T-1} \gamma_t 
    \left\Vert {\rvx}^{(i)}_{0,k} - \Hat{\rvx}^{(i)}_{0,t} \right\Vert^2
    \right]
\right] + C
\\ & = %%%%% 2
\mathbb{E}_{\rvx_0, \rvepsilon}
\left[
\sum_k \sum_{i:\eta_{ik}=1}
    \left[
    \frac{\Bar{\alpha}_T}{2} 
    \left\Vert {\rvx}^{(i)}_{0} - \vmu \right\Vert^2
    +
    \sum_{t=1}^{T-1} \gamma_t 
    \left\Vert {\rvx}^{(i)}_{0} - \Hat{\rvx}^{(i)}_{0,t} \right\Vert^2
    \right]
\right] + C
\end{aligned}
\end{equation}
It is obvious that $\mathbb{E}_{\rvx_0}[\Vert \rvx_0 - \vmu^{(i)}\Vert^2] \leq \mathbb{E}_{\rvx_0}[\Vert \rvx_0 - \vmu \Vert^2]$. Thus we focus on proving the minimum of the second term $\mathbb{E}_{\rvx_0,\rvepsilon}[\sum_k\sum_{i:\eta_{ik}=1}\sum_{t=1}\gamma_t \Vert \rvx_0^{(i)} - \Hat{\rvx}^{(i)}_{0,t} \Vert^2]$ based on the decomposed priors is less than that based on the standard priors. The main difference of this term is about $\Hat{\rvx}^{(i)}_{0,t}=f_\vtheta(\rvx_t, t)$ where ${\rvx}^{(i)}_{t} = \Tilde{\rvx}^{(i)}_{t} + \vmu^{(i)}$ in the decomposed case and ${\rvx}^{(i)}_{t} = \Breve{\rvx}^{(i)}_{t} + \vmu$ in the standard case as shown by \cref{eq:appendix_proof_diffuseion_process_decompose} and \cref{eq:appendix_proof_diffuseion_process_standard}. 

Under a reasonable assumption that $\sum_i\vmu^{(i)} = \vmu = \vzero$, this term in the decomposed and standard case can be derived as \cref{eq:elbo_second_term_decomp} and \cref{eq:elbo_second_term_standard}, respectively. 

\begin{equation}
\begin{split}
\label{eq:elbo_second_term_standard}
& \mathbb{E}_{\rvx_0, \rvepsilon}
\left[
\sum_k \sum_{i:\eta_{ik}=1}
    \sum_{t=1}^{T-1} \gamma_t 
    \left\Vert {\rvx}^{(i)}_{0} - \Hat{\rvx}^{(i)}_{0,t} \right\Vert^2
\right]
\\ = & %%%%% 1
\mathbb{E}_{\rvx_0, \rvepsilon} \left[
\sum_{i}
    \sum_{t=1}^{T-1} \gamma_t 
    \left\Vert {\rvx}^{(i)}_{0} - \left[\theta\sum_{j\neq i}(\rvx_t^{(j)}-\rvx_t^{(i)})+ \rvx_t^{(i)}\right] \right\Vert^2
\right]
\\ = & %%%%% 2
\mathbb{E}_{\rvx_0, \rvepsilon} \left[
\sum_{i}
    \sum_{t=1}^{T-1} \gamma_t 
    \left\Vert \Breve{\rvx}^{(i)}_{0}+\vmu - 
    \theta 
        \sum_{j\neq i} \left[ (\Breve{\rvx}_t^{(j)}+\vmu)-(\Breve{\rvx}_t^{(i)}+\vmu)\right] - \Breve{\rvx}^{(i)} -\vmu \right\Vert^2
\right]
\\ = & %%%%% 3
\mathbb{E}_{\rvx_0, \rvepsilon} \left[
\sum_{i}
    \sum_{t=1}^{T-1} \gamma_t 
    \left\Vert \Breve{\rvx}^{(i)}_{0} - 
    \theta 
        \sum_{j\neq i} \left[ \Breve{\rvx}_t^{(j)} - \Breve{\rvx}_t^{(i)} \right] - \Breve{\rvx}^{(i)} \right\Vert^2
\right]
\\ = & %%%%% 4
\begin{aligned}[t]
\mathbb{E}_{\rvx_0, \rvepsilon} \Bigg[
\sum_{i} 
    \sum_{t=1}^{T-1} \gamma_t 
    & \bigg\Vert \Breve{\rvx}^{(i)}_{0} - \bigg[  
    \sqrt{\Bar{\alpha}_t} \theta \sum_{j\neq i} \Breve{\rvx}_0^{(j)} + \sqrt{1 - \Bar{\alpha}_t} \theta \sum_{j\neq i} \rvepsilon_t^{(j)}
        -(N-1)\sqrt{\Bar{\alpha}_t}  \theta \Breve{\rvx}_0^{(i)} 
        -(N-1)\sqrt{1 - \Bar{\alpha}_t}  \theta \rvepsilon_t^{(i)} 
        \bigg] 
    \\ &  - \left(
    \sqrt{\Bar{\alpha}_t} \Breve{\rvx}_0^{(i)} 
    + \sqrt{1 - \Bar{\alpha}_t} \rvepsilon_t^{(i)} 
    \right) \bigg\Vert^2
    \Bigg]
    \end{aligned}
\\ = & %%%%% 5
\begin{aligned}[t]
\mathbb{E}_{\rvx_0, \rvepsilon} \Bigg[
\sum_{i} 
    \sum_{t=1}^{T-1} \gamma_t 
    \bigg\Vert & 
    (\mI + (N-1)\sqrt{\Bar{\alpha}_t}\theta-\sqrt{\Bar{\alpha}_t}\mI)\Breve{\rvx}_0^{(i)}
    + \sqrt{1-\Bar{\alpha}_t}((N-1)\theta - \mI) \rvepsilon_t^{(i)}
    \\ & - \sqrt{\Bar{\alpha}_t} \theta \sum_{j\neq i} \Breve{\rvx}_0^{(j)}
    - \sqrt{1-\Bar{\alpha}_t} \theta \sum_{j\neq i} \rvepsilon_t^{(j)}
    \bigg\Vert^2
\Bigg]
\end{aligned}
\\ = & %%%%% 6
    \sum_{t=1}^{T-1} \gamma_t 
    \sum_d \Bigg[
    \begin{aligned}[t]
        & \big[ N^2(N-1) +\blue{\sqrt{\Bar{\alpha}_t} [N+(N-1)^2] \sum_i \mu_{d}^{(i)2}}
        \Big] \red{\theta_d^2}
        \\ & + \bigg[ 2N(N-1)(\sqrt{\Bar{\alpha}_t}-1)+\blue{\sqrt{\Bar{\alpha}_t}(1-\sqrt{\Bar{\alpha}_t}) (2N-1) \sum_i \mu_{d}^{(i)2}}\Big] \red{\theta_d}
        \\ & + \bigg[ 
        N(1-\sqrt{\Bar{\alpha}_t})^2 + N(1-{\Bar{\alpha}_t}) + \blue{(1-\sqrt{\Bar{\alpha}_t})^2 \sum_i \mu_{d}^{(i)2}}
        \bigg] 
        \Bigg]
    \end{aligned}
\end{split}
\end{equation}
\begin{equation}
\begin{split}
\label{eq:elbo_second_term_decomp}
& \mathbb{E}_{\rvx_0, \rvepsilon}
\left[
\sum_k \sum_{i:\eta_{ik}=1}
    \sum_{t=1}^{T-1} \gamma_t 
    \left\Vert {\rvx}^{(i)}_{0} - \Hat{\rvx}^{(i)}_{0,t} \right\Vert^2
\right]
\\ = & %%%%% 1
\mathbb{E}_{\rvx_0, \rvepsilon} \left[
\sum_{i}
    \sum_{t=1}^{T-1} \gamma_t 
    \left\Vert {\rvx}^{(i)}_{0} - \left[\theta\sum_{j\neq i}(\rvx_t^{(j)}-\rvx_t^{(i)})+ \rvx_t^{(i)}\right] \right\Vert^2
\right]
\\ = & %%%%% 2
\mathbb{E}_{\rvx_0, \rvepsilon} \left[
\sum_{i}
    \sum_{t=1}^{T-1} \gamma_t 
    \left\Vert \Breve{\rvx}^{(i)}_{0}+\vmu^{(i)} - 
    \theta 
        \sum_{j\neq i} \left[ (\Breve{\rvx}_t^{(j)}+\vmu^{(i)})-(\Breve{\rvx}_t^{(i)}+\vmu^{(j)})\right] - \Breve{\rvx}^{(i)} -\vmu^{(i)} \right\Vert^2
\right]
\\ = & %%%%% 3
\mathbb{E}_{\rvx_0, \rvepsilon} \left[
\sum_{i}
    \sum_{t=1}^{T-1} \gamma_t 
    \left\Vert \Tilde{\rvx}^{(i)}_{0} - 
    \theta 
        \sum_{j\neq i} \left[ \Tilde{\rvx}_t^{(j)} - \Tilde{\rvx}_t^{(i)} \right] - \Tilde{\rvx}^{(i)} \right\Vert^2
\right]
\\ = & %%%%% 4
\begin{aligned}[t]
\mathbb{E}_{\rvx_0, \rvepsilon} \Bigg[
\sum_{i} 
    \sum_{t=1}^{T-1} \gamma_t 
    & \bigg\Vert \Tilde{\rvx}^{(i)}_{0} - \bigg[  
    \sqrt{\Bar{\alpha}_t} \theta \sum_{j\neq i} \Tilde{\rvx}_0^{(j)} + \sqrt{1 - \Bar{\alpha}_t} \theta \sum_{j\neq i} \rvepsilon_t^{(j)}
        -(N-1)\sqrt{\Bar{\alpha}_t}  \theta \Tilde{\rvx}_0^{(i)} 
        -(N-1)\sqrt{1 - \Bar{\alpha}_t}  \theta \rvepsilon_t^{(i)} 
        \bigg] 
    \\ &  - \left(
    \sqrt{\Bar{\alpha}_t} \Tilde{\rvx}_0^{(i)} 
    + \sqrt{1 - \Bar{\alpha}_t} \rvepsilon_t^{(i)} 
    \right) \bigg\Vert^2
    \Bigg]
    \end{aligned}
\\ = & %%%%% 5
\begin{aligned}[t]
\mathbb{E}_{\rvx_0, \rvepsilon} \Bigg[
\sum_{i} 
    \sum_{t=1}^{T-1} \gamma_t 
    \bigg\Vert & 
    (\mI + (N-1)\sqrt{\Bar{\alpha}_t}\theta-\sqrt{\Bar{\alpha}_t}\mI)\Tilde{\rvx}_0^{(i)}
    + \sqrt{1-\Bar{\alpha}_t}((N-1)\theta - \mI) \rvepsilon_t^{(i)}
    \\ & - \sqrt{\Bar{\alpha}_t} \theta \sum_{j\neq i} \Tilde{\rvx}_0^{(j)}
    - \sqrt{1-\Bar{\alpha}_t} \theta \sum_{j\neq i} \rvepsilon_t^{(j)}
    \bigg\Vert^2
\Bigg]
\end{aligned}
\\ = & %%%%% 6
    \sum_{t=1}^{T-1} \gamma_t 
    \sum_d \Bigg[
    \begin{aligned}[t]
        & \Big[ N^2(N-1) 
        \Big] \red{\theta_d^2}
        + \bigg[ 2N(N-1)(\sqrt{\Bar{\alpha}_t}-1)\Big] \red{\theta_d}+ \bigg[ 
        N(1-\sqrt{\Bar{\alpha}_t})^2 + N(1-{\Bar{\alpha}_t}) \bigg] 
        \Bigg]
    \end{aligned}
\end{split}
\end{equation}

The difference between the final derivation of \cref{eq:elbo_second_term_decomp} and \cref{eq:elbo_second_term_standard} are highlighted in \blue{blue}. Thus the minimum along the dimension $d$ over the parameter $\red{\theta_d}$ can be expressed in the same format as follows:
\begin{equation}
\begin{aligned}
&  N \sum_t \gamma_t [(1-\sqrt{\Bar{\alpha}_t})^2 + (1-{\Bar{\alpha}_t})] + \blue{\sum_t \gamma_t 
 (1-\sqrt{\Bar{\alpha}_t})^2 \sum_i \mu_{d}^{(i)2}}
\\ & - 
\frac{
\bigg[ 
2N(N-1)\sum_t\gamma_t(\sqrt{\Bar{\alpha}_t}-1) 
+ \blue{(2N-1) \sum_t\gamma_t \sqrt{\Bar{\alpha}_t} (1- \sqrt{\Bar{\alpha}_t} ) \sum_i \mu_{d}^{(i)2}}
\bigg]^2
}{
4\bigg[ N^2(N-1) 
\sum_t \gamma_t 
+\blue{ [N+(N-1)^2] \sum_t \gamma_t {\Bar{\alpha}_t} \sum_i \mu_{d}^{(i)2} }
\bigg]
}
\end{aligned}
\end{equation}
The above formula can be expressed in the form $f(x)=ax+b-\frac{(ex+f)^2}{cx+d}$ where $x=\sum_i \mu_{d}^{(i)2}$ and the constants $a,b,c,d,e,f$ are all positive in our setting. $f(x)$ is strictly monotone increasing function on $[0,+\infty)$ when $ac>e^2$. Thus our proof is done when the following inequation holds:
\begin{equation}
\label{eq:proof_inequality}
[N+(N-1)^2] \sum_t \gamma_t 
  (1-\sqrt{\Bar{\alpha}_t})^2 
  \sum_t \gamma_t \Bar{\alpha}_t > 
  \Big[(2N-1)\sum_t \gamma_t \sqrt{\Bar{\alpha}_t} (1- \sqrt{\Bar{\alpha}_t} ) \Big]^2
\end{equation}
Obviously, $N+(N-1)^2>(2N-1)^2$. According to Cauchy–Schwarz inequality $\left(\sum_i u_i v_i\right)^2 \leq \left(\sum_i u_i \right)^2 \left(\sum_i v_i\right)^2$, $\sum_t \gamma_t  (1-\sqrt{\Bar{\alpha}_t})^2 \sum_t \Bar{\alpha}_t = \left( \sum_t (\sqrt{\gamma_t}(1-\sqrt{\Bar{\alpha}_t}))^2 \right)\cdot \left( 
\sum_t (\sqrt{\gamma_t\Bar{\alpha}_t})^2
\right) \geq \left( \sum_t (\sqrt{\gamma_t}(1-\sqrt{\Bar{\alpha}_t})) (\sqrt{\gamma_t\Bar{\alpha}_t}) \right)^2 = \left( \sum_t \gamma_t \sqrt{\Bar{\alpha}_t} (1-\sqrt{\Bar{\alpha}_t})  \right)^2 $ also holds. Thus we reach the conclusion $\min_{\theta} -\text{ELBO}_{\text{decomp}} < \min_{\theta} -\text{ELBO}_{\text{standard}}$.

For the aspect (b), we refer to the proof in \citet{lee2021priorgrad} which shows the tighter ELBO with prior $\gN(\vmu, \mSigma)$ than $\gN(\vzero, \mI)$ under some assumptions when $\mSigma$ aligns with the covariance of data distribution. Because the decomposed priors used in the training phase are obtained by Maximum Likelihood Estimation (MLE), they are supposed to have better alignment with data distribution.
\section{Derivation of Validity Guidance}
\label{sec:guidance_derivation}
In this section, we will show the derivation of arms-scaffold drift and clash drift in detail.

The additional drift term that promotes the connection between the arms and scaffold is derived as follows:
% We assert the existence of connections between all the arms and scaffold if the following inequality holds for $n=1:|\mathcal{A}|$,
% \begin{equation}
%     \rho_{\min} \leq \min_{i\in \mathcal{A}_n, j\in \mathcal{S}} \Vert x^{(i)} - x^{(j)} \Vert_2 \leq \rho_{\max}
% \end{equation}
% where $\rho_{\min}$ and $\rho_{\max}$ are hyperparameters approximately representing the range of a bond length and set to $1.2\AA$ and $1.9\AA$ respectively in practice.
\begin{equation}
\begin{split}
\label{eq:arms_scaffold_drift_derivation}
& \nabla_{\vx_t} \log{P(\{ \rho_{\min} \leq \min_{i\in \mathcal{A}_n, j\in \mathcal{S}} \Vert \vx_t^{(i)} - \vx_t^{(j)} \Vert_2 \leq \rho_{\max}, n=1:|\mathcal{A}| \}|\vx_t)} 
\\ %%% 1
= & \nabla_{\vx_t} \sum_{n=1}^{|\mathcal{A}|} \log{P(\{ \rho_{\min} \leq \min_{i\in \mathcal{A}_n, j\in \mathcal{S}} \Vert \vx_t^{(i)} - \vx_t^{(j)} \Vert_2 \leq \rho_{\max}\}|\vx_t)} 
\\ %%%% 2
= &  \sum_{n=1}^{|\mathcal{A}|} \frac{\nabla_{\vx_t} 
\left[
P(\{ -\min_{i\in \mathcal{A}_n, j\in \mathcal{S}} \Vert \vx_t^{(i)} - \vx_t^{(j)} \Vert_2 \leq -\rho_{\min}\}|\vx_t)
\cdot 
P(\{ \min_{i\in \mathcal{A}_n, j\in \mathcal{S}} \Vert \vx_t^{(i)} - \vx_t^{(j)} \Vert_2 \leq \rho_{\max}\}|\vx_t)
\right]}
{P(\{ \rho_{\min} \leq \min_{i\in \mathcal{A}_n, j\in \mathcal{S}} \Vert \vx_t^{(i)} - \vx_t^{(j)} \Vert_2 \leq \rho_{\max}\}|\vx_t)}
\\ %%%% 3
= & \sum_{n=1}^{|\mathcal{A}|}
\zeta_1 \nabla_{\vx_t} P(\{ \min_{i\in \mathcal{A}_n, j\in \mathcal{S}} \Vert \vx_t^{(i)} - \vx_t^{(j)} \Vert_2 \leq \rho_{\max}\}|\vx_t)
+
\zeta_2
\nabla_{\vx_t} P(\{ -\min_{i\in \mathcal{A}_n, j\in \mathcal{S}} \Vert \vx_t^{(i)} - \vx_t^{(j)} \Vert_2 \leq -\rho_{\min}\}|\vx_t)
\\ %%%% 4
= & \sum_{n=1}^{|\mathcal{A}|}
\zeta_1 \mathbb{E}[ \nabla_{\vx_t}\mathbb{I} (-\min_{i\in \mathcal{A}_n, j\in \mathcal{S}} \Vert \vx_t^{(i)} - \vx_t^{(j)} \Vert_2 \leq -\rho_{\min})|\vx_t]
+
\zeta_2
\nabla_{\vx_t}\mathbb{E}[  \mathbb{I} (-\min_{i\in \mathcal{A}_n, j\in \mathcal{S}} \Vert \vx_t^{(i)} - \vx_t^{(j)} \Vert_2 \leq -\rho_{\min})|\vx_t] 
\end{split}
\end{equation}
where $\zeta_1=1/P(\{ -\min_{i\in \mathcal{A}_n, j\in \mathcal{S}} \Vert \vx_t^{(i)} - \vx_t^{(j)} \Vert_2 \leq -\rho_{\min}\}|\vx_t)$ and $\zeta_2=1/P(\{ \min_{i\in \mathcal{A}_n, j\in \mathcal{S}} \Vert \vx_t^{(i)} - \vx_t^{(j)} \Vert_2 \leq \rho_{\max}\}|\vx_t)$.

 The additional drift term that guides our model to generate molecules outside the protein surface is derived as follows:
\begin{equation}
\begin{split}
\label{eq:clash_drift_derivation}
    & \nabla_{\vx_t} \log{P(\{ S(\vx^{(i)}_t)>\gamma, \forall{i}\}|\vx_t)} 
    \\
    = & \frac{\nabla_{\vx_t}P(\{ S(\vx^{(i)}_t)>\gamma, \forall{i}\}|\vx_t) }{P(\{ S(\vx^{(i)}_t)>\gamma, \forall{i}\}|\vx_t)}
    % \\
    % = & \frac{\nabla_{\vx_t}\mathbb{E}[ \sum_{i=1}^{N_M} \mathbb{I} ( S(\vx^{(i)}_t)>\gamma)/N_M|\vx_t] }{\mathbb{E}[ \sum_{i=1}^{N_M} \mathbb{I} ( S(\vx^{(i)}_t)>\gamma)/N_M|\vx_t]}
    \\
    = & \zeta_3 \nabla_{\vx_t}\mathbb{E}[ \sum_{i=1}^{N_M} \mathbb{I} ( -S(\vx^{(i)}_t)<-\gamma)/N_M|\vx_t] 
\end{split}
\end{equation}
where $\zeta_3=1/\mathbb{E}[ \sum_{i=1}^{N_M} \mathbb{I} ( -S(\vx^{(i)}_t)<-\gamma)/N_M|\vx_t]$.

Due to the discontinuity of the indicator function that is incompatible with the gradient operator, we use $\xi-\max{(0,\xi-y)}$ as a surrogate of $\mathbb{I}(y<\xi)$ in \cref{eq:arms_scaffold_drift_derivation} and \cref{eq:clash_drift_derivation}. Although $\zeta_1, \zeta_2, \zeta_3$ are dependent on $\vx_t$, we find setting them as constant still works well. With these two approximations, we can derive the arms-scaffold drift and clash drift as \cref{eq:arms_scaffold_drift_final} and \cref{eq:clash_drift_final} respectively.

\section{Implementation Details}
\label{sec:app:model_details}

\subsection{Featurization}
We represent each protein atom with the following features: one-hot element indicator (H, C, N, O, S, Se), one-hot amino acid type indicator (20 dimension), one-dim flag indicating whether the atom is a backbone atom, and one-hot arm/scaffold region indicator. If the distance between the protein atom and any arm prior center is within 10 \AA, the protein atom will be labeled as belonging to an arm region and otherwise a scaffold region.

The ligand atom is represented with following features: one-hot element indicator (C, N, O, F, P, S, Cl) and one-hot arm/scaffold indicator. Note that the partition of arms and scaffold is predetermined. Thus, the decomposition feature only serves as the input of the model but will not be involved in the network's prediction. 

We build two graphs for message passing in the protein-ligand complex:  a $k$-nearest neighbors graph upon ligand atoms and protein atoms (we choose $k=32$ in all experiments) and a fully-connected graph upon ligand atoms. As Sec. \ref{sec:bond_diff} mentioned, we only need to predict a subset of edges even though the message passing can be performed on all edges. In the knn graph, the edge features are the outer products of distance embedding and edge type. The distance embedding is obtained by expanding distance with radial basis functions located at 20 centers between 0 \AA and 10 \AA. The edge type is a 4-dim one-hot vector indicating the edge is between ligand atoms, protein atoms, ligand-protein atoms or protein-ligand atoms. In the ligand graph, the ligand bond is represented with a one-hot bond type vector (non-bond, single, double, triple, aromatic), an additional feature indicating whether or not two ligand atoms are from the same arm/scaffold prior.

\subsection{Model Details}
Our neural network is mainly composed of three types of layers: atom update layer, bond update layer, and position update layer. In each layer, we apply graph attention to aggregate the message of each node/edge. The key/value/query embedding is obtained through a 2-layer MLP with LayerNorm and ReLU activation. Stacking these three layers as a block, our model consists of 6 blocks with \texttt{hidden\_dim=128} and \texttt{n\_heads=16}. 

We set the number of diffusion steps as 1000. For this diffusion noise schedule, we choose to use a sigmoid $\beta$ schedule with $\beta_1 = \texttt{1e-7}$ and $\beta_T = \texttt{2e-3}$ for atom coordinates, and a cosine $\beta$ schedule suggested in \citet{nichol2021improved} with $s=0.01$ for atom types and bond types.

% \paragraph{Atom Update Layer} This layer takes as input the atom's hidden state $\rvh$, edge features $\rve_\rvf$, and edge index. The change of $\rvh$ is computed based on the graph attention:
% \begin{equation}
% \begin{aligned}
%     \mathbf{K} &= \text{MLP}_1 ([\rve_\rvf, \rvh_i, \rvh_j]) \\
%     \mathbf{V} &= \text{MLP}_2 ([\rve_\rvf, \rvh_i, \rvh_j]) \\ 
%     \mathbf{Q} &= \text{MLP}_3 ([\rve_\rvf, \rvh_i]) \\
%     \Delta \rvh &= \text{MultiHeadAttention}(\mathbf{Q}, \mathbf{K}, \mathbf{V})
% \end{aligned}
% \end{equation}

\subsection{Training Details}
The model is trained via gradient descent method Adam \cite{kingma2014adam} with \texttt{init\_learning\_rate=0.001}, \texttt{betas=(0.95, 0.999)}, \texttt{batch\_size=4} and \texttt{clip\_gradient\_norm=8}. To balance the scales of different losses, we multiply a factor $\alpha=100$ on the atom type loss and bond type loss. During the training phase, we add a small Gaussian noise with a standard deviation of 0.1 to protein atom coordinates as data augmentation. We also schedule to decay the learning rate exponentially with a factor of 0.6 and a minimum learning rate of 1e-6. The learning rate is decayed if there is no improvement for the validation loss in 10 consecutive evaluations. The evaluation is performed for every 2000 training steps. We trained our model on one NVIDIA GeForce GTX A100 GPU, and it could converge within 36 hours and 300k steps.

\section{Additional Results}
\begin{table*}[t]
    % \vspace{-4mm}
    \centering
    \caption{Summary of different properties of reference molecules and molecules generated by multiple variants of our model and TargetDiff for comparison.  ($\uparrow$) / ($\downarrow$) denotes a larger / smaller number is better.
    }
    \begin{adjustbox}{width=1\textwidth}
    \renewcommand{\arraystretch}{1.2}
\begin{tabular}{l|cc|cc|cc|cc|cc|cc|c}
\toprule
% \diagbox{Model}{Metric} 
\multirow{2}{*}{Methods} & \multicolumn{2}{c|}{Vina Score ($\downarrow$)} & \multicolumn{2}{c|}{Vina Min ($\downarrow$)} & \multicolumn{2}{c|}{Vina Dock ($\downarrow$)} & \multicolumn{2}{c|}{High Affinity ($\uparrow$)} & \multicolumn{2}{c|}{QED ($\uparrow$)}   & \multicolumn{2}{c|}{SA ($\uparrow$)} & Success Rate ($\uparrow$) \\
 & Avg. & Med. & Avg. & Med. & Avg. & Med. & Avg. & Med. & Avg. & Med. & Avg. & Med. & Avg. \\
\midrule
Reference   & -6.36 & -6.46 & -6.71 & -6.49 & -7.45 & -7.26 & -  & - & 0.48 & 0.47 & 0.73 & 0.74 & 25.0\%   \\
\midrule

TargetDiff  & -5.47 & -6.30 & -6.64 & -6.83 & -7.80 & -7.91 & 58.1\% & 59.1\% & 0.48 & 0.48 & 0.58 & 0.58 & 10.5\% \\

\method - Atom - Ref Prior & -6.11 & -6.16 & -6.68 & -6.58 & -7.53 & -7.59 &  59.8\% & 63.0\%  & 0.48 & 0.48 & 0.61 & 0.61 & 11.2\% \\

\method - Atom - Pocket Prior & -5.72 & -7.49 & -7.66 & -8.33 & -9.08 & -9.32 & 79.0\% & 96.0\% & 0.41 & 0.38 & 0.52 & 0.52 & 11.4 \% \\

\method - Atom - Opt Prior & -6.01 & -7.34 & -7.65 & -8.01 & -8.93 & -9.02 & 74.5\% & 88.4\%  & 0.37 & 0.34 & 0.50 & 0.50 & 5.8 \% \\

\method - Bond - Ref Prior & -5.17 & -5.25 & -6.03 & -5.98 & -7.10 & -7.14 & 48.9\% & 45.6\% & 0.51 & 0.51 & 0.66 & 0.65 & 15.4\% \\

\method - Bond - Pocket Prior & -5.69 & -6.06 & -7.14 & -7.19 & -8.50 & -8.55 & 69.2\% & 81.6\% & 0.41 & 0.39 & 0.59 & 0.59 & 19.8\% \\

\method - Bond - Opt Prior  & -5.67 & -6.04 & -7.04 & -7.09 & -8.39 & -8.43 & 64.4\% & 71.0\% & 0.45 & 0.43 & 0.61 & 0.60 & 24.5\% \\

\bottomrule
\end{tabular}
\renewcommand{\arraystretch}{1}

    \end{adjustbox}\label{tab:app_mol_prop}
    \vspace{-4mm}
\end{table*}

\subsection{Full Evaluation Results}
\label{sec:full_eval_appendix}
In Table \ref{tab:app_mol_prop}, we show the results of multiple variants of our models. All of them leverage decomposed priors and validity guidance to improve the sampling quality. The differences lie in whether to include bond diffusion and which kind of prior is used. 
\textit{Ref Prior} is estimated from the reference molecule with a Gaussian distribution through maximum likelihood estimation. \textit{Pocket Prior} estimates the prior center as illustrated in Appendix \ref{sec:prior_generation} and uses a neural classifier to estimate the number of ligand atoms and prior standard deviation. \textit{Opt Prior} is a mixture of them which uses Ref Prior if the reference ligand could pass the Success criteria (QED $> 0.25$, SA $> 0.59$, Vina Dock $< -8.18$). It can be seen that bond diffusion consistently has a positive effect on QED and SA. Both bond diffusion and Opt prior have a better balance in the 2D structure rationality and binding affinity, which leads to a higher success rate.

\subsection{Time Complexity}

\begin{table*}[t]
    % \vspace{-4mm}
    \centering
    \caption{Training time of different models.
    }
    % \begin{adjustbox}{width=1\textwidth}
    % \renewcommand{\arraystretch}{1.2}
\begin{tabular}{cccc}
\toprule
Model &	Time(s) / Step	& Total \#Steps & Total Time (hrs) \\
\midrule
AR &	0.15 &	1.5 M &	62.5 \\
Pocket2Mol &	0.55	& 475 K	 & 72.6 \\
TargetDiff & 0.30 &	300 K &	25.0 \\
Ours &	0.50 &	300 K & 41.7 \\
\bottomrule
\end{tabular}
% \renewcommand{\arraystretch}{1}
    % \end{adjustbox}
    \label{tab:time_comp}
    \vspace{-4mm}
\end{table*}

For the training efficiency, we summarize the running time per step and the total running time in \cref{tab:time_comp}. The increase in training time is mainly due to the introduction of bond diffusion, which makes the network more complex. The decomposed prior has a negligible impact on the training time.

For the sampling efficiency, AR, Pocket2Mol, GraphBP, and TargetDiff use 7785s, 2544s, 105s, and 3428s for generating 100 valid molecules on average separately. It takes DecompDiff 5570s / 6189s on average without / with validity guidance. Similarly, the decomposed prior has a negligible impact on the sampling time. Bond diffusion results in 1.62x sampling time compared to TargetDiff, and validity guidance makes the sampling time slightly increase by 10\% further.

\section{Examples of Generated Ligands}
\label{sec:examples_appendix}
\begin{figure*}[ht]
\begin{center}
\centerline{\includegraphics[width=1.0\textwidth]{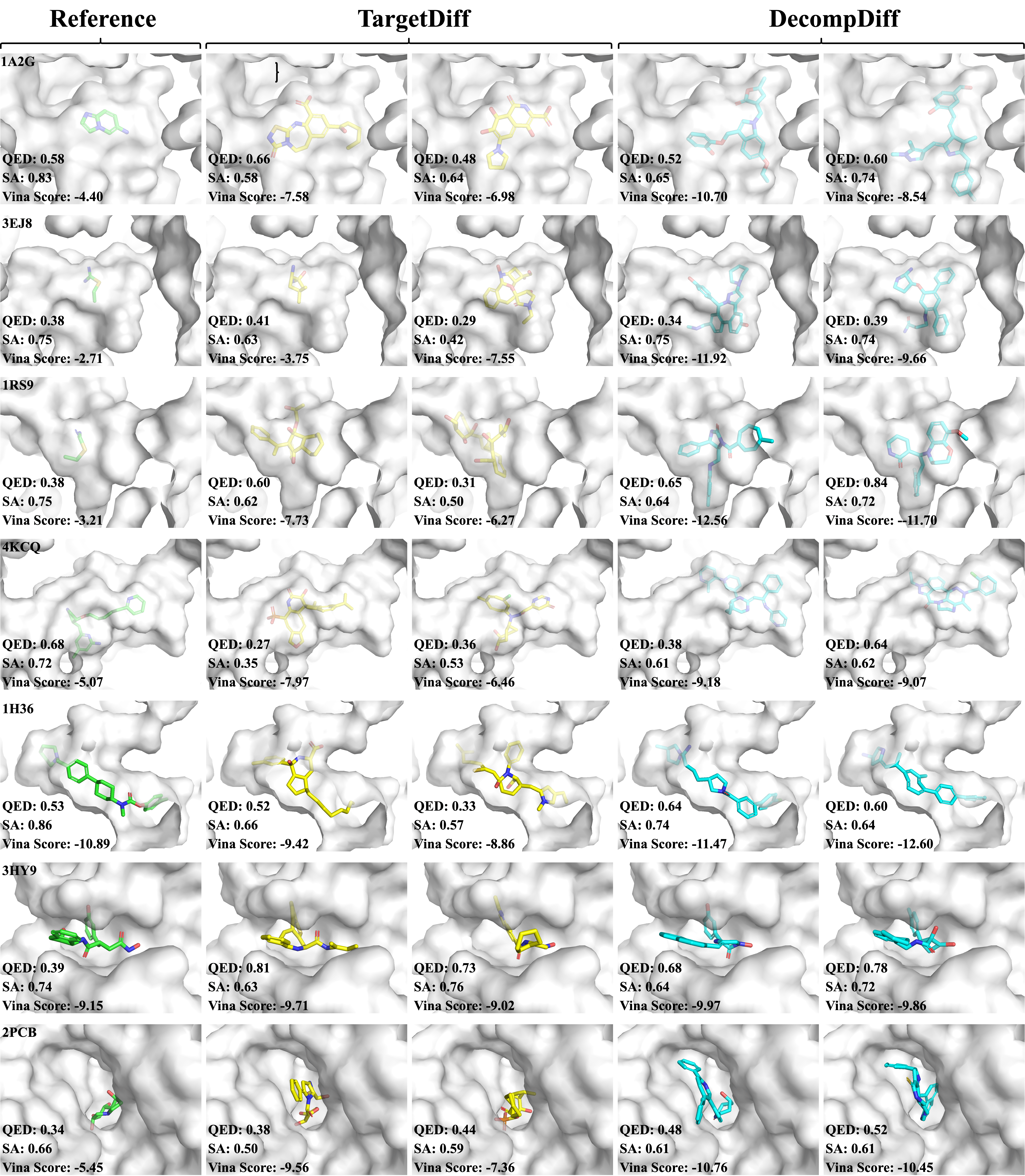}}
\caption{Examples of generated ligands. Carbon atoms in reference ligands, ligands generated by TargetDiff, and our model are visualized in green, yellow, and cyan respectively. Each row corresponds to a protein.}
\label{fig:examples_appendix}
\vspace{-6mm}
\end{center}
\end{figure*}

In \cref{fig:examples_appendix}, we visualize reference ligands and ligands generated by TargetDiff \citep{guan20233d} and our model. As the visualization shows, the ligands generated by our model can occupy more space in the concave target binding site due to the design of decomposed prior. Bond diffusion and validity guidance can promote the quality of generated ligands. Thus the ligands generated by our model can achieve better Vina Scores while keeping reasonable QED and SA.

%%%%%%%%%%%%%%%%%%%%%%%%%%%%%%%%%%%%%%%%%%%%%%%%%%%%%%%%%%%%%%%%%%%%%%%%%%%%%%%
%%%%%%%%%%%%%%%%%%%%%%%%%%%%%%%%%%%%%%%%%%%%%%%%%%%%%%%%%%%%%%%%%%%%%%%%%%%%%%%

\end{document}